ELSEVIER

# Orthogonal parallel MCMC methods for sampling and optimization

L. Martino⋆, V. Elvira†, D. Luengo‡, J. Corander◇, F. Louzada⋆

⋆ Institute of Mathematical Sciences and Computing, Universidade de São Paulo, São Carlos (Brazil).
◇ Dep. of Mathematics and Statistics, University of Helsinki, Helsinki (Finland).
† Dep. of Signal Theory and Communic., Universidad Carlos III de Madrid, Leganés (Spain).
‡ Dep. of Circuits and Systems Engineering, Universidad Politécnica de Madrid, Madrid (Spain).

**Abstract**

Monte Carlo (MC) methods are widely used for Bayesian inference and optimization in statistics, signal processing and machine learning. A well-known class of MC methods are Markov Chain Monte Carlo (MCMC) algorithms. In order to foster better exploration of the state space, specially in high-dimensional applications, several schemes employing multiple parallel MCMC chains have been recently introduced. In this work, we describe a novel parallel interacting MCMC scheme, called *orthogonal MCMC* (O-MCMC), where a set of "vertical" parallel MCMC chains share information using some "horizontal" MCMC techniques working on the entire population of current states. More specifically, the vertical chains are led by random-walk proposals, whereas the horizontal MCMC techniques employ independent proposals, thus allowing an efficient combination of global exploration and local approximation. The interaction is contained in these horizontal iterations. Within the analysis of different implementations of O-MCMC, novel schemes in order to reduce the overall computational cost of parallel multiple try Metropolis (MTM) chains are also presented. Furthermore, a modified version of O-MCMC for optimization is provided by considering parallel simulated annealing (SA) algorithms. Numerical results show the advantages of the proposed sampling scheme in terms of efficiency in the estimation, as well as robustness in terms of independence with respect to initial values and the choice of the parameters.

## 1. Introduction

Monte Carlo (MC) methods are widely employed in different fields for Bayesian inference and stochastic optimization [1, 2, 3, 4]. Markov Chain Monte Carlo (MCMC) methods [5, 6, 4] are well-known MC methodologies to draw random samples and efficiently compute integrals involving a complicated multidimensional target probability density function (pdf), $\pi(\mathbf{x})$ with $\mathbf{x} \in \mathcal{D} \subseteq \mathbb{R}^{d_x}$. MCMC techniques only need to be able to evaluate the target pdf, but the difficulty of diagnosing and speeding up the convergence has driven intensive research efforts in this field. For instance, several adaptive MCMC methods have been developed in order to determine adequately the shape and spread of the proposal density used to generate candidate samples within an MCMC scheme [7, 8, 4, 9]. Nevertheless, guaranteeing the theoretical convergence is still an issue in most of the cases. Moreover, in a single specific (long) run, the generated chain can remain trapped in a local mode and, in this scenario, the adaptation could even slow down the convergence. Thus, in order to speed up the exploration of the state space, and specially to deal with high-dimensional applications, several schemes employing parallel chains have been recently proposed [2, 9], as well as multiple try and interacting schemes [10]. However, the problem is still far from being solved. The interest in the parallel computation can be also originated by other motivations. For instance, several authors have studied the parallelization of MCMC algorithms, which have traditionally been implemented in an iterative non-parallel fashion, in order to reduce their computation time [11, 12].

In this work, we focus on the implementation of parallel MCMC chains in order to foster the exploration of the state space and improve the overall performance. Computational speed up (as result of the parallelization) can be seen as an additional benefit of the proposed approach, but it is not the main goal of the paper. We introduce a novel scheme that considers a population of samples at each iteration, similarly to other population-based techniques [13, 14, 15, 3, 16, 17].[1] More specifically, we present a novel family of parallel MCMC schemes, called orthogonal MCMC (O-MCMC) algorithms, where $N$ different chains are independently run and, at some pre-specified iterations, they exchange information using another MCMC technique applied on the entire population of current states. Assuming that all the MCMC techniques used yield chains converging to the target pdf, the ergodicity of the global scheme is guaranteed: the whole kernel is still valid, since it is obtained as the multiplication of ergodic kernels with the same invariant pdf. Fixing the computational cost, the computing effort can be divided into $N$ parallel processes but, at some iteration, information among the chains is exchanged in order to enhance the overall mixing. Let us remark also that the novel O-MCMC scheme is able to combine efficiently both the random-walk and the independent proposal approaches, as both strategies have advantages and drawbacks. On the one hand, random-walk proposal pdfs are often used when there is no specific information about the target, since this approach turns out to be more explorative than using a fixed proposal. On the other hand, a well-chosen independent proposal density usually provides less correlation among the samples in the generated chain. In the novel method, the parallel "vertical" chains (based on random-walk proposals) move around as "free explorers" roaming the state space, whereas the "horizontal" MCMC technique (applied over the population of current states and based on independent proposals) works as a "park ranger", redirecting "lost explorers" towards the "beaten track" according to the target pdf. Unlike in [19, 20, 21, 22], the exchange of information occurs taking always into account the whole population of current states, instead of applying *crossover* or *exchange* schemes between specific pairs of chains. Tempering of the target pdf is not considered for sampling purposes but it is employed for optimization. Hence, our approach resembles the nonreversible parallel MH algorithms described in [23, 24], where the whole population of states is also updated jointly at the times of interaction, pursuing non-reversibility instead of tempering as a means to accelerate convergence towards posterior mode regions. However, both tempering and crossovers can also be easily implemented within the O-MCMC framework.

Another important contribution of the work is the computational improvement provided by novel parallel implementations of MCMC techniques using multiple candidates at each iteration. We present two novel schemes for parallel Multiple try Metropolis (MTM) chains [10, 25, 26, 27, 28, 29] (and similarly to [12]) in order to reduce the overall computational cost in the same fashion of [11], saving generated samples, target evaluations and multinomial sampling steps. One of them is an extended version, using several candidates, of the Block Independent Metropolis presented in [11]. The ergodicity of both schemes is guaranteed. These novel parallel MTM techniques are employed as horizontal methods in O-MCMC. The corresponding O-MCMC scheme (using a novel parallel MTM method) can also be interpreted as an MTM algorithm employing an adaptive proposal density. This pdf is a mixture of $N$ components: the adaptation of the location parameters of the $N$ components is driven by the vertical parallel chains (note that the outputs of these chains are also used in the estimation). Furthermore, we describe a modified version of O-MCMC for solving optimization problems (where we employ tempering of the target), considering parallel Simulated Annealing algorithms [30, 31, 32] for the vertical movements. Numerical simulations show that O-MCMC exhibits both flexibility and robustness with respect to the initialization and parameterization of the proposals.

It is also important to remark that, in literature, there is a great interested in proposing possible parallel implementation of MCMC algorithms [33, 34, 35, 36, 37], distributing the computing in different parallel processors. However, it is not the goal of this work: we focus on suggesting a novel MCMC scheme which improves the performance w.r.t. other techniques, fixing the number of target density evaluations (similarly to [11, 12]).

The paper is structured as follows. Section 2 summarizes the general framework and the aim of the work. Section

---

[1] A preliminary version of this work has been published in [18]. With respect to that paper, here we propose several novel interacting schemes for exchanging information among the chains, analyze the theoretical basis of the proposed approach and discuss its relationships w.r.t. other techniques, in detail. Different variants are presented in order to reduce the overall computational cost and for applying O-MCMC in optimization problems. Furthermore, we provide more exhaustive numerical simulations.

3 describes the generic O-MCMC scheme, whereas Sections 4 and 5 provide different specific examples of vertical and horizontal movements, respectively. Section 6 discusses the O-MCMC framework for optimization and Section 7 describes the connections with other techniques. Section 8 provides different numerical results. Finally, some concluding remarks are provided in Section 9.

## 2. Bayesian inference problem

In many applications, we aim at inferring a variable of interest given a set of observations or measurements. Let us denote the variable of interest by $\mathbf{x} \in \mathcal{D} \subseteq \mathbb{R}^{d_x}$, and let $\mathbf{y} \in \mathbb{R}^{d_y}$ be the observed data. The posterior pdf is then

$$\bar{\pi}(\mathbf{x}) = p(\mathbf{x}|\mathbf{y}) = \frac{\ell(\mathbf{y}|\mathbf{x})g(\mathbf{x})}{Z(\mathbf{y})}, \tag{1}$$

where $\ell(\mathbf{y}|\mathbf{x})$ is the likelihood function, $g(\mathbf{x})$ is the prior pdf and $Z(\mathbf{y})$ is the model evidence (a.k.a. marginal likelihood). In general, $Z(\mathbf{y})$ is unknown, so we consider the corresponding unnormalized target function,

$$\pi(\mathbf{x}) = \ell(\mathbf{y}|\mathbf{x})g(\mathbf{x}). \tag{2}$$

In general, the analytical study of the posterior density $\bar{\pi}(\mathbf{x}) \propto \pi(\mathbf{x})$ is unfeasible (for instance, integrals involving $\bar{\pi}(\mathbf{x})$ are typically intractable), and numerical approximations are required. Our goal is to approximate efficiently $\bar{\pi}(\mathbf{x})$ employing a cloud of random samples. In general, a direct method for drawing independent samples from $\bar{\pi}(\mathbf{x})$ is not available and alternative approaches (e.g., MCMC algorithms) are needed. The only required assumption is being able to evaluate the unnormalized target function $\pi(\mathbf{x})$.

## 3. O-MCMC algorithms: General outline

Let us consider $N$ parallel vertical chains, $\{\mathbf{x}_{n,t}\}_{n=1}^{N}$ with $t \in \mathbb{N}$, generated by different MCMC techniques with *random-walk* proposal pdfs $q_{n,t}(\mathbf{x}) = q_n(\mathbf{x}|\mathbf{x}_{n,t-1}) = q_n(\mathbf{x} - \mathbf{x}_{n,t-1})$, i.e., $\mathbf{x}_{n,t-1}$ plays the role of a location parameter for the proposal pdf used in the next iteration. Let us denote the population of current states at the $t$-th iteration as

$$\mathcal{P}_t = \{\mathbf{x}_{1,t}, \mathbf{x}_{2,t}, \dots, \mathbf{x}_{N,t}\}.$$

At certain selected iterations, we apply another MCMC technique taking into account the entire population of states $\mathcal{P}_{t-1}$, yielding a new cloud of samples $\mathcal{P}_t$. In this "horizontal" transitions, the different chains share information. The horizontal MCMC technique uses a proposal pdf which is independent from the previous states, unlike the random walk proposals employed in the vertical MCMC chains. The general O-MCMC approach is represented graphically in Figure 1 and summarized below:

1. **Initialization:** Choose the $N$ initial states,

$$\mathcal{P}_0 = \{\mathbf{x}_{1,0}, \mathbf{x}_{2,0}, \dots, \mathbf{x}_{N,0}\},$$

   the total number of iterations, $T$, and three positive integer values $M, T_V, T_H \in \mathbb{N}\backslash\{0\}$ such that $M(T_V + T_H) = T$. Set $t = 1$.
2. **For $m=1,\dots,M$:**
   (a) **Vertical period:** For

$$t = (m-1)(T_V + T_H) + 1, \dots, mT_V + (m-1)T_H,$$

   run $N$ independent MCMC techniques, starting from $\mathbf{x}_{n,t-1} \in \mathcal{P}_{t-1}$, to obtain $\mathbf{x}_{n,t}$ for $n = 1, \dots, N$, i.e., a new population of states $\mathcal{P}_t = \{\mathbf{x}_{1,t}, \mathbf{x}_{2,t}, \dots, \mathbf{x}_{N,t}\}$.
   (b) **Horizontal period:** For

$$t = mT_V + (m-1)T_H + 1, \dots, m(T_V + T_H),$$

   apply an MCMC approach taking into account the entire population $\mathcal{P}_{t-1}$ to generate the next cloud $\mathcal{P}_t$.
3. **Output:** Return the $NT = NM(T_V + T_H)$ samples contained in all the sets $\mathcal{P}_t$, for $t = 1, \dots, T$.

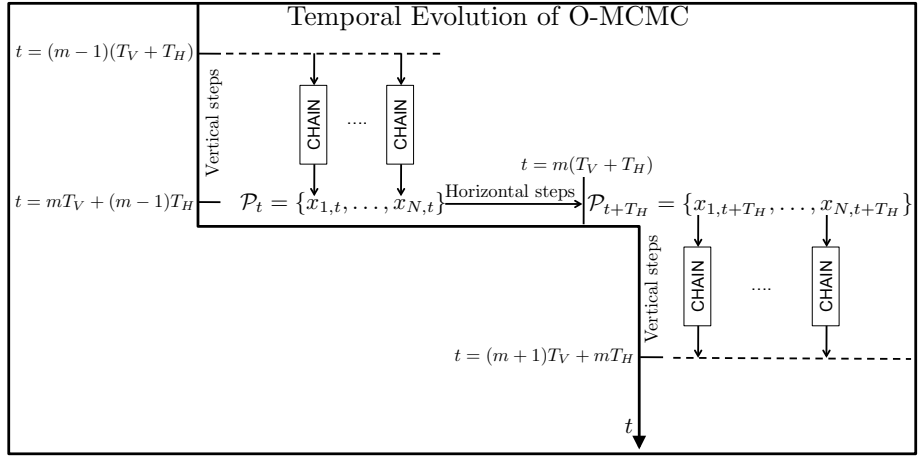

Figure 1. A graphical representation of the O-MCMC approach. After $T_V$ vertical transitions, then $T_H$ horizontal steps are performed.

Table 1. **Notation for O-MCMC.**

| | |
|---|---|
| $N$ | Cardinality of the population, i.e., number of parallel chains. |
| $T_V$ | Iterations per chain at each vertical period. |
| $T_H$ | Iterations per chain at each horizontal period. |
| $M$ | Number of epochs, i.e., cycles of vertical and horizontal periods. |
| $T$ | Total number of iterations of the algorithm, $T = M(T_V + T_H)$. |
| $E_T$ | Total number of evaluation of the target distribution $\bar{\pi}(\mathbf{x}) \propto \pi(\mathbf{x})$. |
| $N \times T$ | Total number of generated samples (states of the chains). |
| $q_n(\mathbf{x}\|\mathbf{x}_{n,t-1})$ | Proposal pdf of the $n$-th chain, for the vertical periods. |
| $\varphi(\mathbf{x})$ | Proposal pdf of the *population approach* for the horizontal periods. |
| $\psi(\mathbf{x}) = \frac{1}{N}\sum_{n=1}^{N}\varphi_n(\mathbf{x}\|\mathbf{x}_{n,t})$ | Proposal pdf of the *mixture-based approach* for the horizontal periods |

In summary, one vertical period contains $T_V$ iterations of the chains, whereas in one horizontal period we have $T_H$ iterations. Hence, given $t = (m-1)(T_V + T_H)$, after one cycle of vertical and horizontal steps we have $t = m(T_V + T_H)$. The total number of cycles (or epochs)[2] is then $M = \frac{T}{T_V + T_H}$. The ergodicity is guaranteed if the vertical and horizontal steps produce ergodic chains with invariant density $\bar{\pi}(\mathbf{x})$ (see Appendix A for further details). Table 1 summarizes the main notation of the paper and the connections of O-MCMC with other techniques are discussed in Section 7. In the following two sections, we introduce several examples of vertical and horizontal movements that lead to different O-MCMC algorithms.

### 3.1. Key observation: burn-in and convergence

In general, several authors have noted that there is not a clear advantage using independent parallel MCMC chains (IPCs) with respect to employing a single longer MCMC chain (fixing the number of evaluation of the target $E_T$) in terms of performance (e.g., see [4, 9, 20, 21, 38]). The reason is that all the shorter parallel chains can remain within their "burn-in" period, thus jeopardizing the global performance, whereas the single longer chain can reach the convergence. Thus, the preference between these two schemes depends on the specific problem [4, 38, 39, 40].

The motivation behind O-MCMC is to take advantage of the aforementioned drawback of the IPCs scheme. Using IPCs we can discover different features of the target pdf in faster way with respect to the use of a single chain, since the different chains will typically concentrate on different areas of the target during the first iterations depending on their initialization. O-MCMC allows the exchange of information among the chains without jeopardizing their ergodicity (see Appendix A). This is particularly useful in multimodal, high-dimensional problems. For instance, using different chains, there are more chances to discover the two modes of the target $\bar{\pi}$ in Figure 2. The horizontal step of O-MCMC allows the communications between the two chains in Figure 2, fostering the identification of the other mode. Indeed, even if some chain is trapped around one mode, O-MCMC can still take advantage of this scenario by redirecting the other chains away from it, and the horizontal stage (which can be interpreted as an alternative to the use of resampling procedures [13, 14, 15]) will eventually cause this chain to move away from that mode. Finally, observe that by employing parallel chains it is possible to apply a diagnosis criterion in order to estimate the "burn-in" period, as already done by other authors [41, 42, 43, 44, 45]. This information can be employed in order to design adaptive strategies, as suggested in [9].

---

[2]One cycle, or epoch, includes one the vertical period and one horizontal period.

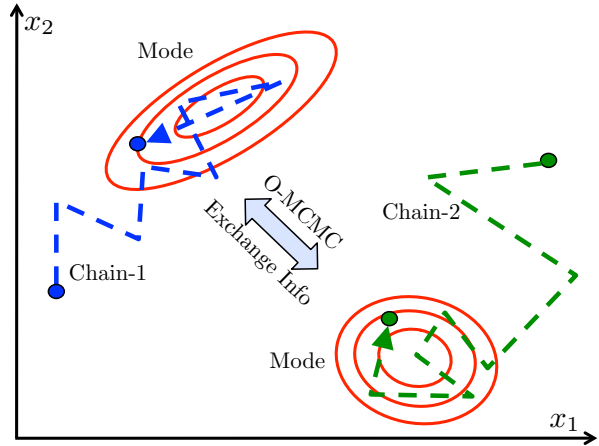

Figure 2. A graphical representation of the key motivation behind the O-MCMC approach. A bivariate, bimodal target pdf $\bar{\pi}(\mathbf{x}) = \bar{\pi}(x_1, x_2)$ is considered (shown its contour plot) and $N = 2$ independent chains are run (their trajectories are depicted with dashed lines), both becoming trapped in a different mode. The horizontal step in O-MCMC fosters the mixing of the two chains by the exchange of information.

## 4. Vertical Movements

In this section, we describe different implementations of the vertical parallel chains. Although it is not strictly necessary, we consider only random walk proposal densities in the vertical chains. The idea is to exploit predominantly the explorative behavior of the independent parallel MCMC methods. Therefore, we only consider proposals of the type $q_n(\mathbf{x}|\mathbf{x}_{n,t-1}) = q_n(\mathbf{x} - \mathbf{x}_{n,t-1})$. In this case, a sample $\mathbf{x}' \sim q_n(\mathbf{x}|\mathbf{x}_{n,t-1})$ can be expressed as

$$\mathbf{x}' = \mathbf{x}_{n,t-1} + \boldsymbol{\xi}_{n,t}, \tag{3}$$

where $\boldsymbol{\xi}_{n,t} \sim q(\boldsymbol{\xi})$. Another more sophisticated possibility is to include the gradient information of the target within the proposal pdf, as suggest in the *Metropolis-Adjusted Langevin Algorithm* (MALA) [46]. In this case, a sample $\mathbf{x}' \sim q_n(\mathbf{x}|\mathbf{x}_{n,t-1})$ becomes

$$\mathbf{x}' = \mathbf{x}_{n,t-1} + \frac{\epsilon}{2}\nabla\log\left[\pi(\mathbf{x}_{n,t-1})\right] + \sqrt{\epsilon}\boldsymbol{\xi}_{n,t}, \tag{4}$$

where $\boldsymbol{\xi}_{n,t} \sim q(\boldsymbol{\xi})$ and $\nabla f(\mathbf{x})$ denotes the gradient of a generic function $f(\mathbf{x})$. This second alternative can be particularly useful in high-dimensional spaces, although it inevitably increases the probability of the chain of becoming trapped in one mode of the target in a multi-modal scenario. Thus, the joint application of $N$ parallel chains appears very appropriate in this scenario, since they can easier reach different modes of the target. Moreover, the application of the O-MCMC scheme facilitates the jumps among the different modes.

Regarding the MCMC algorithm, note that the random walk proposal density $q_n(\mathbf{x}|\mathbf{x}_{n,t-1})$ can be applied within different MCMC kernels. The simplest possibility is using a *Metropolis-Hastings* (MH) algorithm [4]. For each $n = 1, \ldots, N$ and for a given time step $t$, one MH update of the $n$-th chain is obtained as

1. Draw $\mathbf{x}' \sim q_n(\mathbf{x}|\mathbf{x}_{n,t-1})$.

2. Set $\mathbf{x}_{n,t} = \mathbf{x}'$ with probability

$$\alpha_n = \min\left[1, \frac{\pi(\mathbf{x}')q_n(\mathbf{x}_{n,t-1}|\mathbf{x}')}{\pi(\mathbf{x}_{n,t-1})q_n(\mathbf{x}'|\mathbf{x}_{n,t-1})}\right].$$

Otherwise (i.e., with probability $1 - \alpha_n$) set $\mathbf{x}_{n,t} = \mathbf{x}_{n,t-1}$.

Many other alternative schemes can be used instead of MH kernel for the vertical chains. For instance, two particularly appealing alternatives are the Multiple Try Metropolis (MTM) [25, 28] and the Delayed Rejection Metropolis [47] techniques.

## 5. Horizontal Movements

As described above, after each iteration $t$ of the vertical period, the vertical chains return a population $\mathcal{P}_t = \{\mathbf{x}_{1,t}, \dots, \mathbf{x}_{N,t}\}$. When $t = mT_V + (m-1)T_H$, with $m \in \{1, ..., M\}$, i.e., after $T_V$ vertical transitions, then $T_H$ horizontal steps are performed. The purpose of these horizontal MCMC transitions is to exchange information among the $N$ different chains, improving the global mixing. In the following, we consider two different general approaches for sharing the information among the chains:

- In the first one, a population-based MCMC algorithm is applied. The states of the vertical chains contained in $\mathcal{P}_t$ are used as the initial population. Furthermore, the population-based MCMC scheme takes into account all the current population for making decisions about the next population.

- In the second one, named as mixture-based approach, the initial population $\mathcal{P}_t$ is also used for building a suitable density $\psi(\mathbf{x})$. This pdf $\psi$ is employed as proposal by the $N$ parallel MCMC chains for yielding the next populations $\mathcal{P}_{t+1}, \dots, \mathcal{P}_{t+T_H}$. More specifically, in this work we suggest to construct $\psi(\mathbf{x})$ as a mixture of $N$ pdfs, each one centered in $\mathbf{x}_{n,t} \in \mathcal{P}_t$.

In the following we show one specific example of the population-based approach and three different versions of the mixture-based scheme. In all the different cases, for the horizontal movements we consider the use of independent proposal pdfs, unlike for the vertical ones, where we have used of random walk proposals.

### 5.1. Population-based approach

We consider a generalized target density,

$$\bar{\pi}_g(\mathbf{x}_1, \dots, \mathbf{x}_N) \propto \prod_{n=1}^{N} \pi(\mathbf{x}_n), \tag{5}$$

where each marginal, $\pi(\mathbf{x}_n)$ for $n = 1, ..., N$ and $\mathbf{x}_n \in \mathcal{D} \subseteq \mathbb{R}^{d_x}$, coincides with the target pdf in Eq. (2). The idea is that the horizontal MCMC transitions leave invariant the extended target $\bar{\pi}_g$. Namely, after a "burn-in" period, the population $\mathcal{P}_t = \{\mathbf{x}_{1,t}, \dots, \mathbf{x}_{N,t}\}$ is distributed according to $\bar{\pi}_g$. The simplest possible population based scheme consists of employing a standard Metropolis-Hastings (MH) algorithm directly in the extended domain, $\mathcal{D}^N \subseteq \mathbb{R}^{d_x \times N}$, with a target $\bar{\pi}_g$, generating (block) transitions from $\mathcal{P}_t$ to $\mathcal{P}_{t+1}$. However, the probability of accepting a new population in this case becomes negligible as $N$ grows. As an alternative example of a population-based scheme, we consider the *Sample Metropolis-Hastings* (SMH) method [39, Chapter 4]. At each iteration, the underlying idea of SMH is replacing one "bad" sample in the population with a "better" one, according to a certain suitable probability. The new sample, candidate of be incorporated in the population, is generated from and independent proposal pdf $\varphi(\mathbf{x})$. The algorithm is designed so that, after a "burn-in" period $t_b$, the elements in $\mathcal{P}_{t'}$ ($t' > t_b$) are distributed according to $\bar{\pi}_g$ in Eq. (5). Table 2 provides a detailed description of the SMH-based horizontal transitions.

The acceptance probability, $0 \le \alpha \le 1$, depends on the entire population, $\mathbf{x}_{n,t-1}$ for $n = 1, \dots, N$, and the new candidate sample, $\mathbf{x}_{0,t-1}$. At each step, the sample chosen to be replaced is selected according to a probability proportional to the inverse of the corresponding importance weight. The ergodicity can be proved by considering the extended density $\bar{\pi}_g$ as the target pdf (see Appendix D). Let us remark that the difference between $\mathcal{P}_t$ and $\mathcal{P}_{t+1}$ is at most one sample. For this reason, a suggestion for a robust implementation is to set $T_H \ge N$ (so that all the samples are

Table 2. Sample Metropolis-Hastings (SMH) algorithm for horizontal transitions in O-MCMC.

1. For $t = mT_V + (m-1)T_H + 1, \ldots, m(T_V + T_H)$:

    (a) Draw $\mathbf{x}_{0,t-1} \sim \varphi(\mathbf{x})$.

    (b) Choose a "bad" sample $\mathbf{x}_{k,t-1} \in \mathcal{P}_{t-1}$, i.e., select an index $k \in \{1, \ldots, N\}$ with probability proportional to the *inverse* of the importance sampling weights

$$\gamma_k = \frac{\frac{\varphi(\mathbf{x}_{k,t-1})}{\pi(\mathbf{x}_{k,t-1})}}{\sum\limits_{n}^{N} \frac{\varphi(\mathbf{x}_{n,t-1})}{\pi(\mathbf{x}_{n,t-1})}}, \qquad k = 1, \ldots, N.$$

    (c) Accept the new population $\mathcal{P}_t = \{x_{n,t}\}_{n=1}^{N}$, with $\mathbf{x}_{i,t} = \mathbf{x}_{i,t-1}$ for all $i \neq k$ and $\mathbf{x}_{k,t} = \mathbf{x}_{0,t-1}$, with probability

$$\alpha = \frac{\sum_{n=1}^{N} \frac{\varphi(\mathbf{x}_{n,t-1})}{\pi(\mathbf{x}_{n,t-1})}}{\sum\limits_{i=0}^{N} \frac{\varphi(\mathbf{x}_{i,t-1})}{\pi(\mathbf{x}_{i,t-1})} - \min\limits_{0 \leq i \leq N} \frac{\varphi(\mathbf{x}_{i,t-1})}{\pi(\mathbf{x}_{i,t-1})}}. \qquad (6)$$

    Otherwise (i.e., with prob. $1 - \alpha$) set $\mathcal{P}_t = \mathcal{P}_{t-1}$.

potentially replaced), although it is not strictly required as shown in Section 8. Moreover, it can be convenient to use in the estimation only the last population $\mathcal{P}_{t+T_H}$ (excluding the sets among $\mathcal{P}_t$ and $\mathcal{P}_{t+T_H}$, generated in the horizontal step).

Finally, note also that the SMH algorithm becomes the standard MH method for $N = 1$. Hence, for $N = 1$ the specific O-MCMC implementation using SMH consists of applying alternatively two SMH kernels with different types of proposals: a random walk proposal, $q_n(\mathbf{x}|\mathbf{x}_{n,t-1})$, and an independent one, $\varphi(\mathbf{x})$. This a well-known scheme (cf. [4, 39]), which can be seen as a particular case of the O-MCMC family of algorithms.

### 5.2. Mixture-based approach

An alternative approach is defining the following mixture of pdfs, which is updated every $T_V$ vertical transitions,

$$\psi(\mathbf{x}) = \psi_m(\mathbf{x}|\mathcal{P}_t) = \frac{1}{N} \sum_{n=1}^{N} \varphi_n(\mathbf{x}|\mathbf{x}_{n,t}), \qquad (7)$$

where $t = mT_V + (m-1)T_H$, $m = 1, \ldots, M$, and each $\mathbf{x}_{n,t} \in \mathcal{P}_t$ plays the role of the location parameter of the $n$-th component of the mixture, $\varphi_n$. It is important to remark that each component $\varphi_n$ is a density arbitrarily chosen by the user, defined in $\mathcal{D}$ (it can be even a mixture itself). Observe that $\psi(\mathbf{x})$ changes from one horizontal period to the next one (since it depends on the final population of the vertical period), but then it remains fixed within the $T_H$ iterations of each horizontal period. Thus, during the complete O-MCMC run we employ $M$ different mixtures, $\psi_1, \ldots, \psi_M$, one for each horizontal period. However, in order to simplify the notation, we use $\psi(\mathbf{x})$. Figure 3 provides a graphical representation. We employ $\psi(\mathbf{x})$ within $N$ independent MCMC schemes as an independent proposal density, namely independent from the previous state of the chain. The underlying idea is using the information in $\mathcal{P}_t$, with $t = mT_V + (m-1)T_H$, to build a good proposal function for performing $N$ independent MCMC processes. The theoretical motivation is that, after the burn-in periods, the vertical chains have converged to the target, so $\mathbf{x}_{n,t} \sim \bar{\pi}(\mathbf{x})$ for $n = 1, \ldots, N$. Then, $\psi(\mathbf{x})$ in Eq. (7) can be interpreted as a kernel density estimation of $\bar{\pi}$, where $\varphi_n$ play the role of the kernel functions.

#### 5.2.1. Basic schemes

As a first example of this strategy, we consider the application of MH transitions. At each iteration $t = mT_V + (m-1)T_H + 1, \ldots, m(T_V + T_H)$, one sample $\mathbf{x}'$ is generated from $\psi(\mathbf{x})$ and then $N$ different MH tests are performed. The procedure is shown in Table 3 and represented in Figure 4. Alternatively, a different sample $\mathbf{x}'_n$, drawn from $\psi(\mathbf{x})$, can be tested for each chain, as shown in Table 4. Hence, $N$ different samples are drawn at each iteration (instead of only one) but, after building $\psi(\mathbf{x}|\mathcal{P}_t)$, the process could be completely parallelized. The variant in Table 4 provides in general better performance, although at the expense of a increasing computational cost in terms of evaluations of the target and number of generated samples. However, the *block independent MH* methodology [11], proposed in order to reduce the computational effort by recycling generated samples and target evaluations, can be employed. For clarifying that, let us consider for simplicity $T_H = N$. Step 2(a) in Table 3 could be modified by drawing only $N$ independent samples $\mathbf{x}'_1, \ldots, \mathbf{x}'_N$ from $\psi(\mathbf{x})$ and, at each iteration $t$, a different circular permutation of the set $\{\mathbf{x}'_1, \ldots, \mathbf{x}'_N\}$ could be tested in the different $N$ acceptance MH tests[3]. Note that, the scheme in Table 3 yields dependent chains, whereas the algorithm in Table 4 produces independent chains (the interaction, in this case, is only contained in the construction of the mixture $\psi$ at the beginning of the horizontal period). Finally, observe that the procedure in Table 3 presents certain similarities with the Normal Kernel Coupler (NKC) method introduced in [48], thus indicating that NKC-type algorithms can be also employed as alternative population-based approaches.

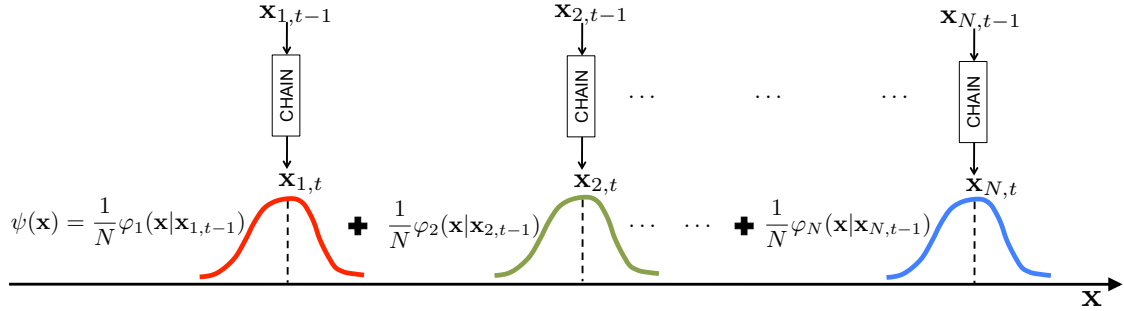

Figure 3. A graphical representation of the mixture-based strategy. The mixture $\psi(\mathbf{x})$ is formed by $N$ components, $\varphi_n(\mathbf{x}|\mathbf{x}_{n,t})$, where $\mathbf{x}_{n,t} \in \mathcal{P}_t$ plays the role of a location parameter. Note that each component $\varphi_n$ can be any kind of density defined in $\mathcal{D}$, even a mixture itself.

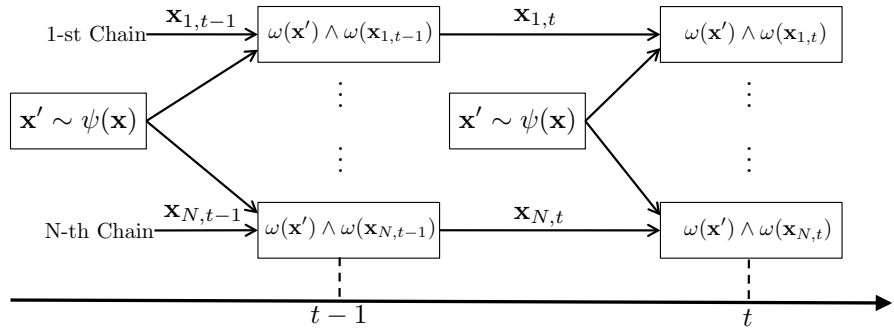

Figure 4. A schematic representation of the basic horizontal scheme described in Table 3. One specific transition of one specific chain is represented with the probability $\alpha_n = \omega(\mathbf{x}') \wedge \omega(\mathbf{x}_{n,t-1})$, where $\omega(\mathbf{x}) = \frac{\pi(\mathbf{x})}{\psi(\mathbf{x})}$, showing the two possible future states at the $t$-th iteration, of the $n$-th chain.

### 5.2.2. Schemes based on multiple candidates

More advanced techniques can also be modified and used as horizontal methods. More specifically, the adaptation to this scenario of multiple try schemes is particularly interesting. For instance, we adjust two special cases[4] of the *Ensemble MCMC* (EnM) [49] and *Multiple Try Metropolis* (MTM) methods [25, 40, 28] to fit them within O-MCMC. Tables 5 and 6 summarize them. Note that standard parallel EnM and MTM chains can be considered. However, we suggest two variants in order to reduce the computational cost. In both cases, $L \geq 1$ different i.i.d. samples, $\mathbf{z}_1, \ldots, \mathbf{z}_L$, are draw from $\psi(\mathbf{x})$. In the parallel Ensemble MCMC (P-EnM) scheme, at each iteration $t$, one resampling step per chain is performed, considering the set of $L+1$ samples $\{\mathbf{z}_1, \ldots, \mathbf{z}_L, \mathbf{x}_{n,t-1}\}$, $n = 1, \ldots, N$, using importance weights. In the parallel MTM (P-MTM) scheme, at each iteration $t$, $N$ resampling steps are performed considering the set of $L$

---

[3]For further clarifications, see the extension of this scheme for a Multiple Try Metropolis method described in Section 5.2.3.

[4]They are special cases of the corresponding algorithms, since an independent proposal pdf $\psi$ is used.

Table 3. Basic mixture scheme for horizontal transitions in O-MCMC.

---

1. Build $\psi(\mathbf{x}) = \psi_m(\mathbf{x}|\mathcal{P}_t)$ as in Eq. (7), where $t = mT_V + (m-1)T_H$.
2. For $t = mT_V + (m-1)T_H + 1, \ldots, m(T_V + T_H)$:
   (a) Draw $\mathbf{x}' \sim \psi(\mathbf{x})$.
   (b) For $n = 1, \ldots, N$:
      i. Set $\mathbf{x}_{n,t} = \mathbf{x}'$, with probability

$$
\begin{aligned}
\alpha_n &= \min\left[1, \frac{\pi(\mathbf{x}')\psi(\mathbf{x}_{n,t-1})}{\pi(\mathbf{x}_{n,t-1})\psi(\mathbf{x}')}\right] \\
&= \omega(\mathbf{x}') \wedge \omega(\mathbf{x}_{n,t-1}),
\end{aligned}
$$

      where $\omega(\mathbf{x}) = \frac{\pi(\mathbf{x})}{\psi(\mathbf{x})}$ and $a \wedge b = \min[a, b]$, for any $a, b \in \mathbb{R}$. Otherwise, set $\mathbf{x}_{n,t} = \mathbf{x}_{n,t-1}$.
   (c) Set $\mathcal{P}_t = \{\mathbf{x}_{1,t}, \ldots, \mathbf{x}_{N,t}\}$.

---

Table 4. Variant of the basic mixture scheme for horizontal transitions in O-MCMC.

---

1. Build $\psi(\mathbf{x}) = \psi_m(\mathbf{x}|\mathcal{P}_t)$ as in Eq. (7), where $t = mT_V + (m-1)T_H$.
2. For $t = mT_V + (m-1)T_H + 1, \ldots, m(T_V + T_H)$:
   (a) For $n = 1, \ldots, N$:
      i. Draw $\mathbf{x}'_n \sim \psi(\mathbf{x})$.
      ii. Set $\mathbf{x}_{n,t} = \mathbf{x}'_n$, with probability

$$
\begin{aligned}
\alpha_n &= \min\left[1, \frac{\pi(\mathbf{x}'_n)\psi(\mathbf{x}_{n,t-1})}{\pi(\mathbf{x}_{n,t-1})\psi(\mathbf{x}'_n)}\right] \\
&= \omega(\mathbf{x}'_n) \wedge \omega(\mathbf{x}_{n,t-1}),
\end{aligned}
$$

      where $\omega(\mathbf{x}) = \frac{\pi(\mathbf{x})}{\psi(\mathbf{x})}$ and $a \wedge b = \min[a, b]$, for any $a, b \in \mathbb{R}$. Otherwise, set $\mathbf{x}_{n,t} = \mathbf{x}_{n,t-1}$.
   (b) Set $\mathcal{P}_t = \{\mathbf{x}_{1,t}, \ldots, \mathbf{x}_{N,t}\}$.

---

candidates $\{\mathbf{z}_1, \ldots, \mathbf{z}_L\}$ and the new possible states are tested (i.e., accepted or not) according to suitable acceptance probabilities $\alpha_n$, $n = 1, \ldots, N$, involving also the previous states $\mathbf{x}_{n,t-1}$. Another alternative and similar technique has been presented in [12], and it is described in Appendix C. This variant uses a non-independent proposal pdf and can be employed as horizontal step.

The ergodicity of both schemes is discussed in Appendix E. The algorithms in Tables 5-6 are obtained by a rearrangement of the basic schemes in [49, 25, 40] in order to generate, at each iteration $t$, $N$ new states for the $N$ independent parallel chains. The new states of the $N$ chains are selected by filtering the same set of candidates $\{\mathbf{z}_1, \ldots, \mathbf{z}_L\}$, drawn from the same independent proposal pdf $\psi$. Note that, with respect to a standard parallel approach, they require less evaluations of the target pdf: at each iteration, the algorithms in Tables 5-6 require $L$ new evaluations of the target instead of the $NL$ target evaluations required by a standard parallel approach. For further explanations, see Appendix E.1.1 and Figure 10. With $L = 1$, the algorithm in Table 5 coincides with the application of $N$ parallel MH methods with Barker's acceptance rule [50]. The algorithm in Table 6 with $L = 1$ coincides with the scheme presented in Table 3. Although any $L \geq 1$ can be employed, a number of tries $L \geq N$ is suggested. Note that another important difference with respect to the standard parallel implementation is that the generated chains are no longer independent.

### 5.2.3. Block Independent Multiple Try Metropolis algorithm

Previously, we have pointed out that with the scheme in Table 6 only $L$ evaluations of the target are required at each iteration, instead of $NL$ as in the standard parallel approach. The proposed scheme in Table 6 can also be modified in the same fashion of the block independent MH method [11], in order to reduce the number of multinomial sampling steps, without jeopardizing the ergodicity of the parallel chains. We remark that the corresponding technique, called

Table 5. Parallel Ensemble MCMC (P-EnM) scheme for horizontal transitions in O-MCMC.

---

1. Build $\psi(\mathbf{x}) = \psi_m(\mathbf{x}|\mathcal{P}_t)$ as in Eq. (7), where $t = mT_V + (m-1)T_H$.
2. For $t = mT_V + (m-1)T_H + 1, \ldots, m(T_V + T_H)$:
   (a) Draw $L$ i.i.d. candidates $\mathbf{z}_1, \ldots, \mathbf{z}_L \sim \psi(\mathbf{x})$.
   (b) For $n = 1, \ldots, N$:
      i. Set $\mathbf{x}_{n,t} = \mathbf{z}_k \in \{\mathbf{z}_1, \ldots, \mathbf{z}_L\}$, i.e., select $k \in \{1, \ldots, L\}$, with probability
      $$\alpha_k = \frac{\frac{\pi(\mathbf{z}_k)}{\psi(\mathbf{z}_k)}}{\sum_{\ell=1}^{L} \frac{\pi(\mathbf{z}_\ell)}{\psi(\mathbf{z}_\ell)} + \frac{\pi(\mathbf{x}_{n,t-1})}{\psi(\mathbf{x}_{n,t-1})}} \quad k = 1, \ldots, L, \tag{8}$$
      or set $\mathbf{x}_{n,t} = \mathbf{x}_{n,t-1}$ with probability
      $$\alpha_{L+1} = 1 - \sum_{k=1}^{L} \alpha_k = \frac{\frac{\pi(\mathbf{x}_{n,t-1})}{\psi(\mathbf{x}_{n,t-1})}}{\sum_{\ell=1}^{L} \frac{\pi(\mathbf{z}_\ell)}{\psi(\mathbf{z}_\ell)} + \frac{\pi(\mathbf{x}_{n,t-1})}{\psi(\mathbf{x}_{n,t-1})}}, \tag{9}$$
      i.e., resample $L$ times the set $\{\mathbf{z}_1, \ldots, \mathbf{z}_L, \mathbf{x}_{n,t-1}\}$ according to the weights $\alpha_k$ for $k = 1, \ldots, L+1$.
      ii. Set $\mathcal{P}_t = \{\mathbf{x}_{1,t}, \ldots, \mathbf{x}_{N,t}\}$.

---

Table 6. Parallel Multiple Try Metropolis (P-MTM) scheme for horizontal transitions in O-MCMC.

---

1. Build $\psi(\mathbf{x}) = \psi_m(\mathbf{x}|\mathcal{P}_t)$ as in Eq. (7), where $t = mT_V + (m-1)T_H$.
2. For $t = mT_V + (m-1)T_H + 1, \ldots, m(T_V + T_H)$:
   (a) Draw $L$ i.i.d. candidates $\mathbf{z}_1, \ldots, \mathbf{z}_L \sim \psi(\mathbf{x})$.
   (b) Draw $N$ independent samples $\{\mathbf{z}_{k_1}, \ldots, \mathbf{z}_{k_N}\}$ such that $\mathbf{z}_{k_n} \in \{\mathbf{z}_1, \ldots, \mathbf{z}_L\}$, i.e., select $k_n \in \{1, \ldots, L\}$ for $n = 1, \ldots, N$, with probability
   $$\beta_{k_n} = \frac{\frac{\pi(\mathbf{z}_{k_n})}{\psi(\mathbf{z}_{k_n})}}{\sum_{\ell=1}^{L} \frac{\pi(\mathbf{z}_\ell)}{\psi(\mathbf{z}_\ell)}}. \tag{10}$$
   Namely, resample $N$ times the samples in the set $\{\mathbf{z}_1, \ldots, \mathbf{z}_L\}$ with probability $\beta_k$, $k = 1, \ldots, L$.
   (c) For $n = 1, \ldots, N$:
      i. Set $\mathbf{x}_{n,t} = \mathbf{z}_{k_n}$ with probability
      $$\alpha_n = \min\left[1, \frac{\sum_{\ell=1}^{L} \frac{\pi(\mathbf{z}_\ell)}{\psi(\mathbf{z}_\ell)}}{\sum_{\ell=1}^{L} \frac{\pi(\mathbf{z}_\ell)}{\psi(\mathbf{z}_\ell)} - \frac{\pi(\mathbf{z}_{k_n})}{\psi(\mathbf{z}_{k_n})} + \frac{\pi(\mathbf{x}_{n,t-1})}{\psi(\mathbf{x}_{n,t-1})}}\right]. \tag{11}$$
      Otherwise, set $\mathbf{x}_{n,t} = \mathbf{x}_{n,t-1}$ (with probability $1 - \alpha_n$).
   (d) Set $\mathcal{P}_t = \{\mathbf{x}_{1,t}, \ldots, \mathbf{x}_{N,t}\}$.

---

*Block Independent Multiple Try Metropolis* (BI-MTM), can always be employed when $N$ parallel independent MTMs are applied (even outside the O-MCMC scheme) in order to reduce the overall computational cost. Let us assume that the value $N$ is such that the number of total transitions of one chain, $T_H$, can be divided in $B = \frac{T_H}{N} \in \mathbb{N}$ blocks. The idea is based on using $N$ circular permutations of the resampled set $\{\mathbf{z}_{k_1}, \ldots, \mathbf{z}_{k_N}\}$, i.e.,

$$
\begin{aligned}
\mathcal{V}_1 &= \{\mathbf{v}_{1,1} = \mathbf{z}_{k_1}, \ldots, \mathbf{v}_{N-1,1} = \mathbf{z}_{k_{N-1}}, \mathbf{v}_{N,1} = \mathbf{z}_{k_N}\}, \\
\mathcal{V}_2 &= \{\mathbf{v}_{1,2} = \mathbf{z}_{k_N}, \ldots, \mathbf{v}_{N-1,2} = \mathbf{z}_{k_{N-2}}, \mathbf{v}_{N,2} = \mathbf{z}_{k_{N-1}}\}, \\
&\vdots \\
\mathcal{V}_N &= \{\mathbf{v}_{1,N} = \mathbf{z}_{k_2}, \ldots, \mathbf{v}_{N-1,N} = \mathbf{z}_{k_N}, \mathbf{v}_{N,N} = \mathbf{z}_{k_1}\},
\end{aligned}
\tag{12}
$$

where each set $\mathcal{V}_n$ denotes one the $N$ possible circular permutations of $\{\mathbf{z}_{k_1}, \ldots, \mathbf{z}_{k_N}\}$. In order to preserve the ergodicity, each $\mathbf{z}_{k_j}$ is drawn from a different set of tries $\mathcal{S}_j = \{\mathbf{z}_1^{(j)}, \ldots, \mathbf{z}_L^{(j)}\}$. More specifically, before a *block* of $N$ iterations, $NL$ tries are drawn from $\psi(\mathbf{x})$, yielding $N$ different sets, $\mathcal{S}_j = \{\mathbf{z}_1^{(j)}, \ldots, \mathbf{z}_L^{(j)}\}$ for $j = 1, \ldots, N$, each one containing $L$ elements. Then, one sample $\mathbf{z}_{k_j}$ is resampled from each $\mathcal{S}_j$ with probability proportional to the corresponding importance weight, and the circular permutations in Eq. (12) are created considering $\{\mathbf{z}_{k_1}, \ldots, \mathbf{z}_{k_N}\}$. The complete BI-MTM algorithm is detailed in Table 13 and further considerations are provided in Appendix E. In Table 13, we have denoted the acceptance probability as $\alpha_n(\mathbf{x}_{n,t-1}, \mathbf{v}_{n,j})$ to remark the two possible future states of the $n$-th chain at the $t$-th iteration. Figure 5 depicts a schematic sketch of the different steps of one block within the BI-MTM algorithm. Moreover, Figure 10 provides a graphical comparison among different parallel MTM approaches. BI-MTM requires only $N$ multinomial sampling steps for each block, i.e., $N$ iterations, instead of $N^2$ as P-MTM in Table 6. Moreover, BI-MTM is completely parallelizable. Indeed, one could draw $NLT_H$ samples from $\psi(\mathbf{x})$, perform $NT_H$ multinomial sampling steps within $NT_H$ different sets, and then run the $T_H$ parallel iterations of the $N$ chains, i.e., one unique block, using circular permutations of the $NT_H$ resampled tries (previously obtained). The reduction in the computational cost is obtained at the expense of a moderate decrease in performance.

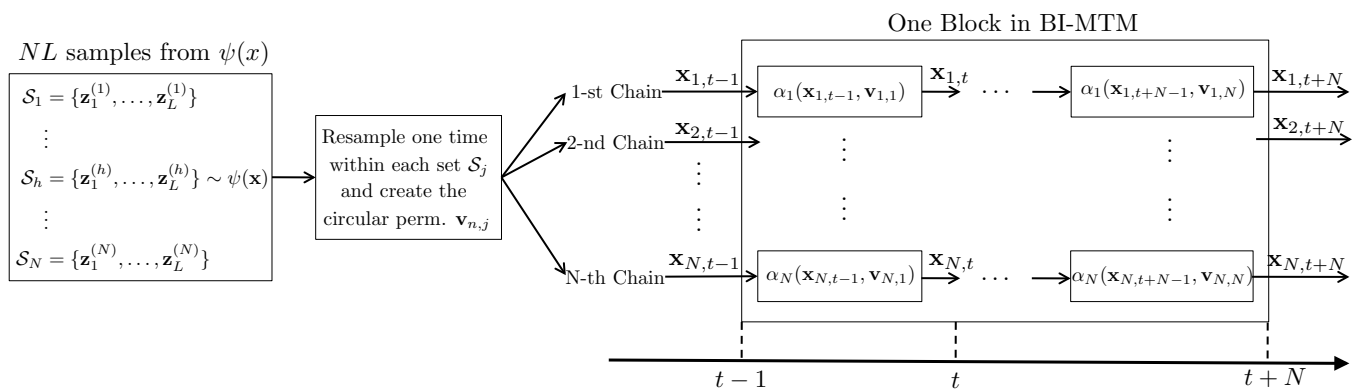

Figure 5. A graphical representation of one block within the BI-MTM technique, described in Table 13. One specific transition of one MTM chain is represented with the probability $\alpha_n(\mathbf{x}_{n,t-1}, \mathbf{v}_{n,j})$, showing the two possible future states at the $t$-th iteration, of the $n$-th chain. One block is formed by $N$ transitions.

### 5.3. Computational cost

In general, the most costly steps are those requiring the evaluation of the target pdf, especially for complex models or a large number of data. The number of evaluations of the target, in one horizontal period, are $E_H = T_H$ for SMH in Table 2, whereas $E_H = LT_H$ in P-EnM and P-MTM (considering, in all cases, only the new evaluations at each iteration, the others can be automatically reused). Using SMH, $T_H$ multinomial sampling steps are performed, each one over a population of $N$ samples. In P-EnM and P-MTM, $NT_H$ multinomial sampling steps are required (with $N > 1$), each one over a set of $L$ samples. The total number of evaluations of the target, $E_T = M(E_V + E_H)$, including the vertical transitions, is $E_T = M(NT_V + T_H)$ when the SMH is employed in the horizontal steps, or $E_T = M(NT_V + LT_H)$ when P-EnM and P-MTM are employed. Furthermore, in BI-MTM, we have again $E_T = M(NT_V + LT_H)$, but only $T_H$ multinomial sampling steps. Note also that in a standard parallel multiple try approach we would have $E_H = NLT_H$ evaluations of the target and $NT_H$ multinomial sampling steps, each one over a set of $L$ samples. Finally, we remark that, using SMH, we perform one acceptance test in each step, i.e., $T_H$ in one horizontal period. Using a multiple candidates scheme, we employ $NT_H$ acceptance test in one horizontal period. All these considerations are summarized in Table 7. For further details and observations, see Appendix E.1.1.

### 5.4. Communication cost

Let us consider briefly now the development of a truly parallel implementation of O-MCMC that can be distributed across different processors/machines. The vertical steps of O-MCMC can be clearly parallelized. However, O-MCMC needs a fusion center in order to perform the horizontal steps. In the mixture-based approach, i.e., O-MCMC-PMTM, the whole population of current states $\mathcal{P}_t = \{\mathbf{x}_{n,t}\}_{n=1}^N$ must be transmitted to this fusion center. If the fusion is performed after each vertical iteration, i.e., $T_V = 1$, then some states, $\mathbf{x}_{n,t} \in \mathcal{P}_t$, are likely to remain unchanged from the previous horizontal step, and thus only certain new (possibly high-dimensional) vectors $\mathbf{x}_{n,t}$ have to be transmitted to the fusion center (indeed, the rest of states have been already transmitted to the fusion center in the previous horizontal step). In other cases, quantization and differential transmission strategies may alleviate the communication cost.

Table 7. Computional cost of O-MCMC given different horizontal schemes. Recall that the number of epochs is $M = \frac{T}{T_V + T_H}$.

| Computational features | SMH | P-EnM and P-MTM | BI-MTM | Stand. Parallel MTM |
|---|---|---|---|---|
| $E_H$ | $T_H$ | $LT_H$ | $LT_H$ | $NLT_H$ |
| $E_T = M(E_V + E_H)$ | $M(NT_V + T_H)$ | $M(NT_V + LT_H)$ | $M(NT_V + LT_H)$ | $M(NT_V + NLT_H)$ |
| Total number of multinomial sampling steps | $MT_H$ | $MNT_H$ | $MT_H$ | $MNT_H$ |
| Cardinality of set for the multinomial sampling | $N$ | $L$ | $L$ | $L$ |
| Total number of acceptance tests | $M(NT_V + T_H)$ | $M(NT_V + NT_H)$ | $M(NT_V + NT_H)$ | $M(NT_V + NT_H)$ |

Note that this communication problem also occurs in many other state of the art algorithms, although it can be reduced through a proper design of the algorithm. For instance, in population-based techniques that employ resampling procedures [13, 15], only the scalar importance weights have to be transmitted and, after the resampling stage, the fusion center can simply return the indices of the resampled particles. In our O-MCMC-SMH, we can follow the same strategy, transmitting only the scalar importance weights, as in [13, 15]. After $T_H$ steps of SMH, the fusion center returns the novel states to the corresponding chains, that can be identified simply through an index.

However, in other more sophisticated schemes that construct the importance weights by considering the so-called deterministic mixture approach [3, 14, 51], the entire set $\mathcal{P}_t$ must be transmitted, as in O-MCMC-PMTM. Similarly, the technique proposed in [12] and described in Appendix C, requires the knowledge of the $L$ candidates for the computation of the weights in Eq. (26). Finally, in the MCMCMC (MC$^3$) method [20, 21], the communication cost is reduced w.r.t. O-MCMC by applying exchanges of particles between specific pairs of chains, whereas in the particle island approach [52] local resampling stages (which only require a subset of particles) are usually performed, with a global resampling stage (that requires all the particles) being performed only occasionally. This kind of strategies could be easily incorporated to the O-MCMC framework in order to enhance its distributed implementation.

### 5.5. Joint adaptation of the proposal densities

Let us denote as $\mathbf{C}_n$ and $\mathbf{\Lambda}_n$ the covariance matrices of the vertical and horizontal proposal pdfs, respectively. In order to design an algorithm as robust as possible, we suggest keeping the scale parameters $\mathbf{C}_n$ fixed for the vertical proposal pdfs $q_n(\mathbf{x}|\mathbf{x}_{n,t-1})$, to avoid a loss of diversity within the set of chosen variances. However, if desired, they could be easily adapted as suggested in [9]. On the other hand, we suggest adapting the scale parameters of the horizontal proposal pdfs $\varphi_n$, $n = 1, \ldots, N$, since it is less delicate. Indeed, let us recall that a poor choice of the $\varphi_n$'s entails an increase in the computational cost, but the diversity in the cloud of samples is always preserved. Several strategies have been proposed in [7, 53] and [9], for adapting proposal functions online within MCMC schemes. For the sake of simplicity, we discuss separately the cases of the population-based or the mixture-based approaches.

- *Adaptation within SMH:* in this case, the strategies in [53, 9] are appropriate. Thus, After a training period $T_{train} < T$, *all* the generated samples (i.e., for each $t > T_{train}$ and from all the chains) can be used to adapt the location and scale parameters of the proposal pdf $\varphi(\mathbf{x})$. Namely, denoting $\varphi_t(\mathbf{x}) = \varphi(\mathbf{x}; \boldsymbol{\mu}_t, \mathbf{\Lambda}_t)$, we can use the following approach:

  - If $t \leq T_{train}$: set $\boldsymbol{\mu}_t = \boldsymbol{\mu}_0$, $\mathbf{\Lambda}_t = \mathbf{\Lambda}_0$ (where $\boldsymbol{\mu}_0$ and $\mathbf{\Lambda}_0$ are the initial choices).
  - If $t > T_{train}$: set $\boldsymbol{\mu}_t = \frac{1}{Nt} \sum_{j=1}^{t} \sum_{n=1}^{N} \mathbf{x}_{n,j}$, and $\mathbf{\Lambda}_t = \frac{1}{Nt} \sum_{j=1}^{t} \sum_{n=1}^{N} (\mathbf{x}_{n,j} - \boldsymbol{\mu}_t)(\mathbf{x}_{n,j} - \boldsymbol{\mu}_t)^\top + \mathbf{C}$, where $\mathbf{C}$ is a chosen covariance matrix. The empirical mean and covariance matrix estimators can also be computed recursively [7].

- *Adaptation of the mixture $\psi(\mathbf{x})$:* the methods in Section 5.2 employ a mixture $\psi(\mathbf{x}) = \frac{1}{N} \sum_{n=1}^{N} \varphi_n(\mathbf{x})$. In this case, every component

$$\varphi_{n,t}(\mathbf{x}) = \varphi_{n,t}(\mathbf{x}; \boldsymbol{\mu}_t, \mathbf{\Lambda}_t),$$

should be adapted, jointly with the weights of the mixture. A possible (and simple) adaptation scheme is provided in [7], where all the parameters of the mixture are updated online. The method in [7] can be easily reformulated for a framework with parallel chains. In this case, the states of the parallel chains are divided into $N$ different clusters according to the Euclidean distance between them and location parameters of the $N$ components in the mixture $\psi(\mathbf{x})$. Then, new centroids (i.e., location parameters), covariance matrices and weights are updated according to the mean, covariance and cardinality of each cluster, respectively.

Finally we remark that, within the O-MCMC framework, it is straightforward to apply the well-known diagnostic criteria in[41, 42, 43, 44, 45] in order to estimate the "burn-in" period, and hence to help the design of the adaptation scheme [9].

## 6. O-MCMC for optimization

The O-MCMC schemes can be easily modified converting them in stochastic optimization algorithms. Indeed, it is possible to replace the $N$ vertical MH chains with $N$ parallel *simulated annealing* (SA) methods [30, 31]. Let us denote as $\gamma_{n,t} \in (0, +\infty)$ a finite scale parameter that is a decreasing function of $t$, approaching zero for $t \to +\infty$, i.e.,

$$\begin{cases} \gamma_{n,t} \geq \gamma_{n,t+1} \geq \ldots \geq \gamma_{n,t+\tau} > 0, \\ \lim_{t \to +\infty} \gamma_{n,t} = 0, \end{cases} \tag{13}$$

for $n = 1, \ldots, N$. Moreover, for the sake of simplicity, we consider symmetric proposal functions $q_n(\mathbf{y}|\mathbf{x}) = q_n(\mathbf{x}|\mathbf{y})$. Then, one transition of the $n$-th SA is described below:

1. Draw $\mathbf{x}' \sim q_n(\mathbf{x}|\mathbf{x}_{n,t-1})$.
2. Set $\mathbf{x}_{n,t} = \mathbf{x}'$ with probability

$$\alpha_n = \min\left[1, \frac{[\pi(\mathbf{x}')]^{\frac{1}{\gamma_{n,t}}}}{[\pi(\mathbf{x}_{n,t-1})]^{\frac{1}{\gamma_{n,t}}}}\right] = \min\left[1, \left(\frac{\pi(\mathbf{x}')}{\pi(\mathbf{x}_{n,t-1})}\right)^{\frac{1}{\gamma_{n,t}}}\right].$$

Otherwise, i.e., with probability $1 - \alpha_n$, set $\mathbf{x}_{n,t} = \mathbf{x}_{n,t-1}$.

Note that, with respect to the MH algorithm, we have replaced the target $\pi(\mathbf{x}) > 0$ with a modified target $[\pi(\mathbf{x})]^{\frac{1}{\gamma_{n,t}}} > 0$, with modes that become sharper and narrower when we reduce the scale parameter $\gamma_{n,t}$. Note also that the movements such $\pi(\mathbf{x}') > \pi(\mathbf{x}_{n,t-1})$ are always accepted, whereas movements leading to $\pi(\mathbf{x}') < \pi(\mathbf{x}_{n,t-1})$ are accepted with probability

$$P_d = \left(\frac{\pi(\mathbf{x}')}{\pi(\mathbf{x}_{n,t-1})}\right)^{\frac{1}{\gamma_{n,t}}} \in (0, 1].$$

This probability $P_d \to 0$ vanishes to zero as $\gamma_{n,t} \to 0$ (guaranteeing the convergence to the global maximum when $t \to +\infty$). In the same fashion, the modified target $[\pi(\mathbf{x})]^{\frac{1}{\gamma_{n,t}}}$ is employed in the horizontal transitions of the mixture-based approach, whereas for the horizontal steps of the population-based approach we consider the modified extended target,

$$\bar{\pi}_g(\mathbf{x}_1, \ldots, \mathbf{x}_N) \propto \prod_{n=1}^{N} [\pi(\mathbf{x}_n)]^{\frac{1}{\gamma_{n,t}}}, \tag{14}$$

so that all the presented schemes, previously described, can be automatically applied. Several possible decreasing functions $\gamma_{n,t}$ have been suggested in [30, 32, 31]. For sampling and optimization purpose, instead of using an artificial sequence of auxiliary parameters $\gamma_{n,t}, \gamma_{n,t+1}, \ldots, \gamma_{n,t+\tau}$, an alternative is to use the so called "data point tempered" techniques [54] where a sequence of $P$ posteriors, $\pi_1(\mathbf{x}), \pi_2(\mathbf{x}),...,\pi_P(\mathbf{x})$, with an increasing number of data, are considered (typically, for sampling purpose the last one contains all the data, i.e., $\pi_P(\mathbf{x}) = \pi(\mathbf{x})$).

## 7. Relationship with other techniques

First, we recall that in this work we focus on population-based Monte Carlo schemes designed in order to foster the exploration of the state space and improve the overall performance. Note that the techniques shown in Table 5 and 6 are interesting since they involve the use of resampling steps without jeopardizing the ergodicity of the resulting global O-MCMC process. Moreover, the SMH algorithm in Table 2 employs an *inverted* resampling scheme, since a sample in the population is chosen to be replaced with probability proportional to the inverse of its importance weight. Other methodologies in the literature employ a combination of MCMC iterations and resampling steps. An example is sequential Monte Carlo (SMC) sampler for a static scenario [15] described in Appendix B.2. The underlying idea could be interpreted belonging to the O-MCMC philosophy: in these methodologies, the resampling steps are seen as a "horizontal" approach for exchanging information within the population. The resampling procedure generates samples from a particle approximation

$$\hat{\pi}^{(L)}(\mathbf{x}) = \sum_{\ell=1}^{L} \beta_\ell \delta(\mathbf{x} - \mathbf{z}_\ell), \tag{15}$$

of the measure of $\bar{\pi}(\mathbf{x})$, where $\mathbf{z}_\ell \sim \psi(\mathbf{x})$ (or, similarly, $\mathbf{z}_\ell \sim q_\ell(\mathbf{x})$ [51]) and $\beta_\ell$ are defined in Eq. (10) in Table 6, with $\ell = 1, \ldots, L$. The quality of this approximation improves as the number $L$ of samples grows. However, for a finite value of $L$ there exists a discrepancy which can produce problems in the corresponding sampling algorithm. For further details see Appendix B. One important issue is the loss in diversity in the population.

This problem is reduced in O-MCMC, since the ergodicity is ensured in both the vertical and the horizontal movements. This improvement in the performance is obtained at the expense of increasing the computational cost. For instance, let us consider the use of SMH in horizontal transitions. The cloud of samples is not impoverished by the application of SMH, even if a poor choice of the proposal $\varphi(\mathbf{x})$ is made. In the worst case, the newly proposed samples are always discarded and computational time is wasted. In the best case, a proposal located in a low probability region can jump close to a mode of the target. Clearly, in the mixture multiple try approach, it is better to choose $L \geq N$ for fostering the safeguard of the diversity. Moreover, in the mixture approach, the mixture $\psi(\mathbf{x}) = \psi(\mathbf{x}|\mathcal{P}_t)$ is built using the states in $\mathcal{P}_t$ as location parameters, and then it does not change for the next $T_H$ horizontal steps. Thus, the information contained in the states $\{\mathbf{x}_{n,t}\}_{n=1}^{N} \in \mathcal{P}_t$ is employed in the next $T_H$ iterations even if some states are not well-located. For clarifying this point, consider for instance the basic scheme in Table 3. The mixture $\psi(\mathbf{x}) = \psi(\mathbf{x}|\mathcal{P}_t)$ does not change, so the information provided by the population $\mathcal{P}_t = \{\mathbf{x}_{1,t}, \ldots, \mathbf{x}_{N,t}\}$ at the iteration $t$ is still used in the iterations $t + 1, \ldots, t + T_H$. This feature is also the main difference between the scheme in Table 3 and the NKC-type methods [48], where one component of the mixture is relocated after each iteration. Unlike the MCMCMC (MC$^3$) method [19, 20, 21, 22], in O-MCMC the exchange of information occurs taking always into account the whole population of current states, instead of applying exchanges between specific pairs of chains. Similarities with the technique proposed in [12] are discussed in Appendix C.

## 8. Numerical simulations

### 8.1. Multimodal target distribution

In this section, we consider a bivariate multimodal target pdf, which is itself a mixture of 5 Gaussian pdfs, i.e.,

$$\bar{\pi}(\mathbf{x}) = \pi(\mathbf{x}) = \frac{1}{5} \sum_{i=1}^{5} \mathcal{N}(\mathbf{x}; \nu_i, \mathbf{G}_i), \quad \mathbf{x} \in \mathbb{R}^2, \tag{16}$$

with means $\nu_1 = [-10, -10]^\top$, $\nu_2 = [0, 16]^\top$, $\nu_3 = [13, 8]^\top$, $\nu_4 = [-9, 7]^\top$, and $\nu_5 = [14, -14]^\top$, and with covariance matrices

$$\mathbf{G}_1 = [2, 0.6; 0.6, 1], \quad \mathbf{G}_2 = [2, -0.4; -0.4, 2],$$
$$\mathbf{G}_3 = [2, 0.8; 0.8, 2], \quad \mathbf{G}_4 = [3, 0; 0, 0.5], \quad \text{and}$$
$$\mathbf{G}_5 = [2, -0.1; -0.1, 2].$$

Then, the target pdf $\pi(\mathbf{x})$ has 5 different modes. We apply O-MCMC to estimate the expected value $E[\mathbf{X}]$ of $\mathbf{X} \sim \bar{\pi}(\mathbf{x})$ (the true mean is $E[\mathbf{X}] = [1.6, 1.4]^\top$) using different values for the number of parallel chains $N \in \{5, 100, 1000\}$. Furthermore, we choose deliberately a "bad" initialization to test the robustness of the algorithm. Specifically, we set $\mathbf{x}_{n,0} \sim \mathcal{U}([-4, 4] \times [-4, 4])$ for $n = 1, \ldots, N$. This initialization is "bad" in the sense that it does not cover the modes of $\pi(\mathbf{x})$. In all cases, we consider MH vertical kernels, with $q_n(\mathbf{x}|\mathbf{x}_{n,t-1}) = \mathcal{N}(\mathbf{x}; \mathbf{x}_{n,t-1}, \mathbf{C}_n)$ as proposal pdfs, using the same isotropic covariance matrix, $\mathbf{C}_n = \sigma^2 \mathbf{I}_2$, for all $n = 1, \ldots, N$. We test different values of $\sigma \in \{2, 5, 10, 70\}$ to gauge the performance of O-MCMC. In O-MCMC, we consider the application of SMH and P-MTM as horizontal techniques, as described below. In both cases, we adapt the covariance matrices of the proposal pdfs as suggested in Section 5.5.

- **O-MCMC with SMH:** As horizontal proposal, we use again a Gaussian pdf, $\varphi_t(\mathbf{x}) = \mathcal{N}(\mathbf{x}; \boldsymbol{\mu}_t, \boldsymbol{\Lambda}_t)$ where $\boldsymbol{\mu}_t$ and $\boldsymbol{\Lambda}_t$ are adapted online: namely, $\boldsymbol{\mu}_t = \frac{1}{Nt} \sum_{j=1}^{t} \sum_{n=1}^{N} \mathbf{x}_{n,j}$, and $\boldsymbol{\Lambda}_t = \frac{1}{Nt} \sum_{j=1}^{t} \sum_{n=1}^{N} (\mathbf{x}_{n,j} - \boldsymbol{\mu}_t)(\mathbf{x}_{n,j} - \boldsymbol{\mu}_t)^\top + \boldsymbol{\Lambda}_0$, where $\boldsymbol{\mu}_0 = [0, 0]^\top$, $\boldsymbol{\Lambda}_0 = \lambda^2 \mathbf{I}_2$ with $\lambda = 2.5$ As remarked in Section 5.5, this adaptive procedure is quite robust since employs samples generated by different parallel chains [9]. Furthermore, we fix $T = 4000$ and $T_H = T_V$. We test different values of $T_V \in \{1, 100\}$ and, as a consequence, $M = \frac{T}{T_V + T_H} = \frac{T}{2T_V} \in \{20, 2000\}$.[6] Recall that the total number of evaluations of the targets in O-MCMC with SMH is $E_T = M(NT_V + T_H) = \frac{T}{2}(N + 1)$ (see Section 5.3).

- **O-MCMC with P-MTM:** We also test the O-MCMC scheme with P-MTM as horizontal technique. In this case, the (independent) proposal pdf is the mixture $\psi(\mathbf{x}) = \psi_m(\mathbf{x}|\mathcal{P}_t) = \frac{1}{N} \sum_{n=1}^{N} \varphi_n(\mathbf{x}|\mathbf{x}_{n,t}, \boldsymbol{\Lambda}_t)$ with $t = mT_V + (m - 1)T_H$, $\boldsymbol{\Lambda}_t = \frac{1}{Nt} \sum_{j=1}^{t} \sum_{n=1}^{N} (\mathbf{x}_{n,j} - \boldsymbol{\mu}_t)(\mathbf{x}_{n,j} - \boldsymbol{\mu}_t)^\top + \boldsymbol{\Lambda}_0$, where $\boldsymbol{\mu}_t = \frac{1}{Nt} \sum_{j=1}^{t} \sum_{n=1}^{N} \mathbf{x}_{n,j}$, $\boldsymbol{\Lambda}_0 = 4\mathbf{I}_2$ (for all $n = 1, \ldots, N$). We consider different number of tries $L = \{5, 50\}$ and set again $T_V = T_H$. In this case, the number of evolution of the target is $E_T = M(NT_V + LT_H) = \frac{T}{2}(N + L)$ (see Section 5.3).

**Comparison with independent parallel chains (IPCs).** We compare the performance of O-MCMC with the application of IPCs, namely, only vertical independent transitions (also in this case the initial state is chosen randomly for each chain at each run, i.e, $\mathbf{x}_{n,0} \sim \mathcal{U}([-4, 4] \times [-4, 4])$ for $n = 1, \ldots, N$). Therefore, we can infer the benefit of applying the horizontal interaction. For a fair comparison, in IPCs we use the same MH kernels, i.e., with the same proposals $q_n$'s, and we keep fixed the total number of evaluations of the target $E_T$ in both cases, O-MCMC and IPCs. Note that $E_T = NT'$ in IPCs where $N$ is the number of chains and $T'$ the total number of iterations for each one. We test different values of $N$. Tables 8 and 9 show the Mean Square Error (MSE), averaged among the two dimensions, in the estimation of the expected value $E[\mathbf{X}] = [1.6, 1.4]^\top$, averaged over 200 independent runs. O-MCMC with SMH always outperforms IPCs, specially for small $\sigma$ and $N$. O-MCMC shows a much more stable behavior w.r.t. the parameter choice $\sigma$. For large scale parameters ($\sigma \in \{10, 70\}$) and a large number of chains ($N \in \{100, 1000\}$), the MSE of IPCs approaches the MSE of O-MCMC. A possible explanation is that the interaction is particularly useful with small $N$ and a wrong choice of $\sigma$, whereas the use of large number of chains such as $N = 100$ or $N = 1000$ is enough, in this bidimensional example, for obtaining good performance. Moreover, O-MCMC with SMH presents an anomalous behavior when the variance of the vertical proposal pdfs is $\sigma = 2$. In this specific case, i.e., only for $\sigma = 2$, the MSE seems increases with $N$. However, note that O-MCMC provides the lower MSE, in any cases, comparing with the same computational effort $E_T$ (with the exception of O-MCMC with P-MTM and $\sigma = 10$).

**Comparison with a single MCMC chain.** We test a single MH chain, i.e., $N = 1$, with a longer length $T$ of the chain, in order to perform the same number of evaluation of the target $E_T$. Note that, in this case, $E_T = T$. Furthermore, we test the adaptive MH method (A-MH) [53] and the delayed rejection MH method (DR-MH) [47]. For A-MH, we consider 10% of the total iterations as a training period (before adaptive the covariance matrix of the proposal). In DR-MH, we consider at most 3 acceptance test before deciding the next state of the chain. At the $t$-th iteration, in each acceptance test of DR-MH, we use a Gaussian proposal pdf with mean the average between the previous mean value and the point rejected at the previous test (at the first stage, the proposal pdf has the current state $\mathbf{x}_t$ as mean). Since in DR-MH we can have more than one evaluation of the target at each iteration (at most 3), and since we fix the total number of evaluations $E_T$, in general the total number of iterations $T'$ is random and varies at each run. Again we set $\mathbf{x}_0 \sim \mathcal{U}([-4,4] \times [-4,4])$, randomly chosen at each independent run, for each method. The results in Tables 8 and 10 show that in this example the use of parallel chains is more convenient in terms of performance. Namely, IPCs and O-MCMC provide a smaller MSE than a single-longer MCMC chains.

---

[5] We set $T_{train} = T_V$, i.e., the adaptation starts after that the samples of the first vertical period are collected. Thus, before of the first horizontal step, $\varphi_t(\mathbf{x})$ has been already updated. However, an estimation of the "burn-in" period could be used for automatically tuning $T_{train}$ [43].

[6] We use all the generated samples in the estimation without removing any "burn-in" period.

**Comparison with Population Monte Carlo (PMC).** We also compare with the standard PMC technique [13], described in Appendix B. We use $N \in \{100, 500, 2000\}$ and $T = 2000$ for PMC, so that the total number of evaluations of the target is $E_T = NT \in \{2 \cdot 10^5, 10 \cdot 10^5, 40 \cdot 10^5\}$. The proposal pdfs used in PMC are the same that we apply for the vertical chains in O-MCMC, i.e., $q_n(\mathbf{x}|\mathbf{x}_{n,t-1}) = \mathcal{N}(\mathbf{x}; \mathbf{x}_{n,t-1}, \mathbf{C}_n)$ using again the same covariance matrix, $\mathbf{C}_n = \sigma^2 \mathbf{I}_2$, for $n = 1, \ldots, N$ and $\sigma \in \{2, 5, 10, 70\}$ (again $\mathbf{x}_{0,n} \sim \mathcal{U}([-4,4] \times [-4,4])$ for $n = 1, \ldots, N$). We have considered a higher number of $E_T$ for PMC with respect to O-MCMC, since O-MCMC involves several acceptance tests which are not contained in PMC. Thus, in order to provide a comparison as fair as possible, we allow a greater number of evaluations of the target, $E_T$, for PMC. Table 10 shows the MSE (mean of the MSEs of each component) of the O-MCMC schemes and the PMC method, for estimating $E[\mathbf{X}]$. We can see that the O-MCMC schemes, even with less $E_T$, provide lower MSEs with the exception of the cases corresponding to $\sigma = 10$.

**Comparison with an adaptive SMC scheme.** Finally, we compare with the SMC scheme described in Appendix B.2, where $N$ parallel MCMC chains, generating the population $\{\mathbf{x}_{n,t}\}_{n=1}^N$, and the interaction is performed by a resampling step, after drawing $N$ samples $\{\mathbf{x}_{n,t+1}\}_{n=1}^N$, each one from $\mathbf{x}_{n,t+1} \sim \varphi_n(\mathbf{x}|\mathbf{x}_{n,t}, \mathbf{\Lambda}_t)$ (we set $T_V = 1$). The resampling plays a role similar to the orthogonal steps in O-MCMC. Thus, for providing the fairest comparison as possible, we also consider here and adaptive covariance matrix $\mathbf{\Lambda}_t = \frac{1}{Nt} \sum_{j=1}^t \sum_{n=1}^N (\mathbf{x}_{n,j} - \boldsymbol{\mu}_t)(\mathbf{x}_{n,j} - \boldsymbol{\mu}_t)^\top + \mathbf{\Lambda}_0$, where $\boldsymbol{\mu}_t = \frac{1}{Nt} \sum_{j=1}^t \sum_{n=1}^N \mathbf{x}_{n,j}$ and $\mathbf{\Lambda}_0 = 4\mathbf{I}_2$. Also, in this case, we have $\mathbf{x}_{n,0} \sim \mathcal{U}([-4,4] \times [-4,4])$ for $n = 1, \ldots, N$, as for O-MCMC. Note that at each $t$-th iteration, $t = 1, \ldots, T$, the resample-move SMC scheme performs a multinomial resampling with cardinality $N$ and then one step of $N$ parallel MCMC chains. As a consequence, the total number evaluations of the target is $E_T = 2NT$, recalling that we set $T_V = 1$ (see App. B.2). The results shown in Table 10. In general, O-MCMC outperforms SMC, considering a similar number $E_T$ of target evaluations. For further comparison between O-MCMC and SMC see Section 8.3.

Computational times (in seconds) are also provided in Table 11.[7] We can observe that, with a Matlab implementation, O-MCMC is also competitive in terms of computational time. Due to the efficient matrix operations (at least with a Matlab implementation), the use of parallel chains is always more convenient in terms of computational time than the use of a single-longer chain (given a fixed number $E_T$ of target evaluations). However, in a specific scenario, a single chain with a longer run could perform better than shorter parallel chains. In this highly multimodal example the use of IPCs is more appropriate (see Tables 8- 9).

| | O-MCMC with SMH | | | | | | Independent parallel chains (IPCs) | | | Single MH chain | | |
|---|---|---|---|---|---|---|---|---|---|---|---|---|
| $N$ | 5 | | 100 | | 1000 | | 5 | 100 | 1000 | 1 | | |
| $T_V$ | 1 | 100 | 1 | 100 | 1 | 100 | | $T$ | | | $T$ | |
| $\sigma = 2$ | 1.4881 | 2.3649 | 1.7515 | 2.9146 | 5.6803 | 6.1354 | 28.7856 | 8.2925 | 7.3543 | 89.6476 | 87.1911 | 41.6461 |
| $\sigma = 5$ | 1.4989 | 2.1724 | 1.4512 | 1.7089 | 1.3606 | 1.4825 | 13.0602 | 2.2842 | 1.8373 | 47.7092 | 5.8160 | 0.6027 |
| $\sigma = 10$ | 1.1769 | 1.4034 | 0.1062 | 0.1129 | 0.0142 | 0.0139 | 2.4443 | 0.1247 | 0.0128 | 2.6611 | 0.1397 | 0.0274 |
| $\sigma = 70$ | 1.8175 | 2.0730 | 0.3554 | 0.3483 | 0.2866 | 0.2815 | 5.4897 | 0.5469 | 0.3264 | 7.5976 | 0.5103 | 0.3271 |
| $T$ | 4000 | | | | | | 2400 | 2020 | 2002 | $12 \cdot 10^3$ | $2.02 \cdot 10^5$ | $20.02 \cdot 10^5$ |
| $E_T$ | $12 \cdot 10^3$ | | $2.02 \cdot 10^5$ | | $20.02 \cdot 10^5$ | | $12 \cdot 10^3$ | $2.02 \cdot 10^5$ | $20.02 \cdot 10^5$ | $12 \cdot 10^3$ | $2.02 \cdot 10^5$ | $20.02 \cdot 10^5$ |

Table 8. Mean Square Error (MSE) in the estimation of the mean of the target, using O-MCMC with SMH and IPCs, considering different values of $\sigma$ and $T_V$ (recall, we set $T_V = T_H$). The total number of evaluations of the target $E_T$ is the same for O-MCMC (where $E_T = \frac{T}{2}(N+1)$ since $T_V = T_H$) and IPCs (where $E_T = NT$). Note that $E_T = T$ for the single MH chain.

---

[7] In order to provide the computational times, the methods are tested in a Laptop-Mac-Processor 1.7 GHz-8 GB-1600 MHz-DDR3. A preliminary Matlab code of O-MCMC-SMH is also provided in `http://www.mathworks.com/matlabcentral/fileexchange/58207-omcmc-smh?s_tid=srchtitle` (note that this code is not optimized).

| | O-MCMC with P-MTM | | | | IPCs | | Single MH chain | | | Single A-MH | Single DR-MH |
|---|---|---|---|---|---|---|---|---|---|---|---|
| $N$ | 5 | | 50 | | 5 | 50 | 1 | | | 1 | 1 |
| $T_V$ | 1 | | | | $T$ | | $T$ | | | $T$ | $T'$ |
| $L$ | 5 | 50 | 5 | 50 | – | – | – | – | – | – | 3 |
| $\sigma = 2$ | 1.3907 | 1.1421 | 1.3156 | 0.7678 | 17.1352 | 3.5950 | 91.7211 | 87.0583 | 85.4229 | 36.2071 | 80.5092 |
| $\sigma = 5$ | 1.6159 | 0.9074 | 1.4011 | 1.0072 | 12.5791 | 2.3277 | 37.9583 | 10.3945 | 7.1659 | 2.9921 | 6.5127 |
| $\sigma = 10$ | 1.5738 | 0.8634 | 1.0982 | 0.8379 | 0.9403 | 0.1134 | 1.4694 | 0.2697 | 0.1375 | 0.1287 | 0.1385 |
| $T$ | 4000 | | | | 4000 | | $2 \cdot 10^4$ | $11 \cdot 10^4$ | $20 \cdot 10^4$ | $20 \cdot 10^4$ | $T'$: dep. on the amount of rejections at each step |
| $E_T$ | $2 \cdot 10^4$ | $11 \cdot 10^4$ | $11 \cdot 10^4$ | $20 \cdot 10^4$ | $2 \cdot 10^4$ | $20 \cdot 10^4$ | $2 \cdot 10^4$ | $11 \cdot 10^4$ | $20 \cdot 10^4$ | $20 \cdot 10^4$ | $20 \cdot 10^4$ |

Table 9. Mean Square Error (MSE) in the estimation of the mean of the target, using O-MCMC with P-MTM and IPCs, considering different values of $\sigma$ and $T_V$ (recall, we set $T_V = T_H$). The total number of evaluations of the target $E_T$ is the same for O-MCMC (where $E_T = \frac{T}{2}(N + L)$ since $T_V = T_H$) and IPCs (where $E_T = NT$). Note that $E_T = T$ for the single MCMC chains. However, for DR-MH, we have $E_T = T'$ where $T'$ depends on the number of rejections occurred in the specific run. The maximum number of acceptance tests in DR-MH, at the same iteration, is set to 3 (this parameter plays a similar role of the number of tries $L$ in the MTM schemes).

| | O-MCMC with SMH | | | O-MCMC with P-MTM | | PMC | | | adaptive SMC | |
|---|---|---|---|---|---|---|---|---|---|---|
| $N$ | 5 | 100 | 1000 | 5 | 50 | 100 | 500 | 2000 | 100 | 200 |
| $\sigma = 2$ | 1.4881 | 1.7515 | 5.6803 | 1.3907 | 1.3156 | 48.11 | 35.1772 | 28.4326 | 2.1114 | 1.1252 |
| $\sigma = 5$ | 1.4989 | 1.4512 | 1.3606 | 1.6159 | 1.4011 | 2.5998 | 2.3230 | 1.9153 | 2.0800 | 1.1501 |
| $\sigma = 10$ | 1.1769 | 0.1062 | 0.0142 | 1.5738 | 1.0982 | 0.0512 | 0.0141 | 0.0054 | 1.6845 | 0.9873 |
| $\sigma = 70$ | 1.8175 | 0.3554 | 0.2866 | 2.0185 | 1.8019 | 2.3963 | 0.8252 | 0.1161 | 1.8899 | 0.9943 |
| $T$ | 4000 | | | | | 2000 | | | 1000 | |
| $E_T$ | $12 \cdot 10^3$ | $2.02 \cdot 10^5$ | $20.02 \cdot 10^5$ | $2 \cdot 10^4$ | $11 \cdot 10^4$ | $2 \cdot 10^5$ | $10 \cdot 10^5$ | $40 \cdot 10^5$ | $2 \cdot 10^5$ | $4 \cdot 10^5$ |

Table 10. Mean Square Error (MSE) in the estimation of the mean of the target, using O-MCMC ($T_V = T_H = 1$ and $L = 5$ for P-MTM) and the standard PMC method [13]. The total number of evaluations of the target is $E_T = \frac{T}{2}(N+1)$ for O-MCMC-SMH, $E_T = \frac{T}{2}(N+L)$ for O-MCMC-SMH (since $T_V = T_H$ for both), $E_T = NT$ for PMC, and $E_T = 2NT$ for adaptive SMC.

### 8.2. Spectral analysis: estimating the frequencies of a noisy multi-sinusoidal signal

Many problems in science and engineering require dealing with a noisy multi-sinusoidal signal, whose general form is given by

$$y_c(\tau) = A_0 + \sum_{i=1}^{d_x} A_i \cos(2\pi f_i \tau + \phi_i) + r(\tau), \quad \tau \in \mathbb{R},$$

where $A_0$ is a constant term, $d_x$ is the number of sinusoids, $\{A_i\}_{i=1}^{d_x}$ is the set of amplitudes, $\{2\pi f_i\}_{i=1}^{d_x}$ are the frequencies, $\{\phi_i\}_{i=1}^{d_x}$ their phases, and $r(\tau)$ is an additive white Gaussian noise (AWGN) term. The estimation of the parameters of this signal is required by many applications in signal processing [55, 56], in control (where a multi-harmonic disturbance is often encountered in industrial plants) [57, 58] or in digital communications (where multiple narrowband interferers can be roughly modeled as sinusoidal signals) [59, 60]. Let us assume that we have $d_y$ equispaced points from $y_c(\tau)$, obtained discretizing $y_c(\tau)$ with a period $T_s < \frac{\pi}{\max_{1 \le i \le d_x} 2\pi f_i}$ (in order to fulfill the sampling theorem [61]):

$$y[k] = A_0 + \sum_{i=1}^{d_x} A_i \cos(\Omega_i k + \phi_i) + r[k], \quad k = 1, \dots, d_y,$$

where $y[k] = y_c(kT_s)$ for $k = 0, 1, \dots, d_y - 1$, $\Omega_i = 2\pi f_i T_s$ for $i = 1, \dots, d_x$, and $r[k] \sim \mathcal{N}(0, \sigma_w^2)$. Our goal is applying the O-MCMC-type algorithms to provide an accurate estimate of the set of unknown frequencies, $\{\Omega_i\}_{i=1}^{d_x}$ or merely $\{f_i\}_{i=1}^{d_x}$. For keeping the notation of the rest of the work, we define the vector of possible frequencies as $\mathbf{x} \in \mathbb{R}^{d_x}$. Thus, given a fixed considering the hyper-rectangular domain $\mathcal{D} = \left[0, \frac{1}{2}\right]^{d_x}$ (it is straightforward to note the periodicity outside $\mathcal{D}$), and a uniform prior on $\mathcal{D}$, the posterior distribution given $K$ data is $\bar{\pi}(\mathbf{x}) \propto \exp\left(-V(\mathbf{x})\right)$, where

$$V(x_1, \dots, x_{d_x}) = \frac{1}{2\sigma_w^2} \sum_{k=1}^{d_y} \left(y[k] - A_0 - \sum_{i=1}^{d_x} A_i \cos(x_i k + \phi_i)\right)^2 \mathbb{I}_{\mathcal{D}}(\mathbf{x}).$$

| Method | $E_T = 10^5$ | | | | | | |
|---|---|---|---|---|---|---|---|
| | $N = 100$ | $N = 500$ | $N = 1000$ | $N = 2000$ | $N = 3000$ | $N = 5000$ | $N = 10^4$ |
| **O-MCMC-SMH** ($T_V = T_H = 1$) | 8.14 (P4) | 2.16 (P4) | 1.21 (P4) | 0.67 (P3) | 0.54 (P2) | 0.39 (P2) | 0.26 (P2) |
| **O-MCMC-SMH** ($T_V = 10, T_H = 1$) | 2.33 (P3) | 0.53 (P3) | 0.29 (P1) | 0.21 (P1) | 0.12 (P1) | 0.11 (P1) | 0.09 (P1) |
| **PMC** | 1.02 (P2) | 0.52 (P2) | 0.60 (P3) | 0.87 (P4) | 1.22 (P4) | 1.81 (P4) | 3.27 (P4) |
| **Adaptive SMC** | 0.84 (P1) | 0.33 (P1) | 0.37 (P2) | 0.48 (P2) | 0.63 (P3) | 0.93 (P3) | 1.64 (P3) |
| **Single MH chain** ($N = 1$ and $T = E_T = 10^5$) | 83.43 (P5) | | | | | | |
| **Single A-MH chain** ($N = 1$ and $T = E_T = 10^5$) | 83.90 (P6) | | | | | | |
| **Single DR-MH chain** ($N = 1$ and $T = E_T = 10^5$) | 88.26 (P7) | | | | | | |

Table 11. Computational time (sec.) of the different algorithms with a Matlab implementation, as function of $N$ and keeping fixed $E_T = 10^5$. The other parameters of the techniques vary in order to keep fixed $E_T = 10^5$ as $N$ grows. We have also highlighted the ranking as P1, P2, P3,P4, P5, P6, and P7, where P1 is the fastest method and P7 the slowest one.

We have denoted $\mathbb{I}_{\mathcal{D}}(\mathbf{x})$ the indicator function such that $\mathbb{I}_{\mathcal{D}}(\mathbf{x}) = 1$ if $\mathbf{x} \in \mathcal{D}$ and $\mathbb{I}_{\mathcal{D}}(\mathbf{x}) = 0$ if $\mathbf{x} \notin \mathcal{D}$. Moreover, for the sake of simplicity, we have assumed also that $S$ and $\sigma_w^2$ are known. Furthermore, we set $A_0 = 0$, $A_i = A = 1$ and $\phi_i = 0.$ [8] Note that the problem is symmetric with respect to the hyperplane $x_1 = x_2 = \ldots = x_{d_x}$ (and, in general, multimodal). Bidimensional examples of $V(\mathbf{x}) = \log \pi(\mathbf{x})$ are depicted in Figure 6. We apply O-MCMC, comparing with IPCs, in two different types of experiments described briefly below. In all cases, we set $\mathbf{x}_{n,0} \sim \mathcal{U}(\mathcal{D})$ for $n = 1, \ldots, N$, $T_H = T_V = 1$, and consider the proposals $q_n(\mathbf{x}|\mathbf{x}_{n,t-1}) = \mathcal{N}(\mathbf{x}; \mathbf{x}_{n,t-1}, \mathbf{C}_n)$ with $\mathbf{C}_n = \sigma^2 \mathbf{I}_{d_x}$, $n = 1, \ldots, N$, for the vertical chains ($\mathbf{I}_{d_x}$ is the unit matrix of dimension $d_x \times d_x$).

**First experiment.** We set $\mathbf{f} = [f_1 = 0.1, f_2 = 0.3]^\top$ and generate $d_y = 10$ synthetic data from the model. Since in this case, $d_x = 2$ and $\mathcal{D} = \left[0, \frac{1}{2}\right]^2$, it is possible to approximate the posterior with a very thin grid and compute the first 4 non-central moments and, as a consequence, we can compare the performance of different Monte Carlo sampling methods. Then, we test O-MCMC-SMH with the horizontal proposal $\varphi(\mathbf{x}) = \mathcal{N}(\mathbf{x}; \boldsymbol{\mu}, \boldsymbol{\Lambda})$ where $\boldsymbol{\mu} = [0.25, 0.25]^\top$ and $\boldsymbol{\Lambda} = \sigma^2 \mathbf{I}_2$, i.e., uses the same $\sigma$ considered for the vertical chains (recall that $\mathbf{C}_n = \sigma^2 \mathbf{I}_2$). We set the total number of target evaluations $E_T = M(N + 1) \in \{2730, 5450, 10.9 \cdot 10^3\}$. For a fair comparison, we consider $N$ independent parallel chains (IPCs) choosing $T$ such that $E'_T = NT$ is equal to $E_T$, i.e., $E'_T = E_T$ (see Section 5.3). We test different values of $\sigma \in [0.05, 0.5]$ and $N \in \{2, 5, 10\}$. We test the combinations of number of chains $N$ and epochs $M$ ($T$ for IPCs) in order to keep fixed $E_T$. The Relative Error (RE) in the estimation, averaged over 500 independent runs, is shown in Figure 7. We can observe that O-MCMC (solid line) outperforms IPCs (dashed line) providing lower REs. The performance becomes similar as the computational effort $E_T$ grows since the state space in the first experiment, $\mathcal{D} = \left[0, \frac{1}{2}\right]^2$, is small enough for allowing an exhaustive exploration of $\mathcal{D}$ by independent chains.

**Second experiment.** We test O-MCMC in higher dimension considering $d_x = 4$, i.e., $\mathcal{D} = \left[0, \frac{1}{2}\right]^4$. We fix $\mathbf{f} = [f_1 = 0.1, f_2 = 0.2, f_3 = 0.3, f_4 = 0.4]^\top$. In this experiment, we consider an optimization problem for finding the global mode of $\pi$ with $d_y = 30$ observations. With $d_y = 30$ observations, the main mode is enough tight around $\mathbf{f}$, so that we consider $\mathbf{f}$ as true localization of the global mode. For simplicity and for breaking the symmetry, we restrict the search to the simplex contained in $\mathcal{D}$ with vertices at $[0, 0, 0, 0]^\top$, $[\frac{1}{2}, 0, 0, 0]^\top$, $[\frac{1}{2}, \frac{1}{2}, 0, 0]^\top$, $[\frac{1}{2}, \frac{1}{2}, \frac{1}{2}, 0]^\top$ and $[\frac{1}{2}, \frac{1}{2}, \frac{1}{2}, \frac{1}{2}]^\top$. We test O-MCMC-PMTM considering again Gaussian horizontal proposals in the mixture $\psi$, with $\boldsymbol{\Lambda} = \lambda^2 \mathbf{I}_4$ for all $n$. We test $\lambda = 0.1$ and $\lambda = \sigma$ (where $\sigma$ is employed in the covariance matrices $\mathbf{C}_n = \sigma^2 \mathbf{I}_4$ of the vertical chains). Moreover, we test the adaptation of $\boldsymbol{\Lambda}$, i.e., $\boldsymbol{\Lambda}_t = \frac{1}{Nt} \sum_{j=1}^t \sum_{n=1}^N (\mathbf{x}_{n,j} - \boldsymbol{\mu}_t)(\mathbf{x}_{n,j} - \boldsymbol{\mu}_t)^\top + \boldsymbol{\Lambda}_0$, where $\boldsymbol{\mu}_t = \frac{1}{Nt} \sum_{j=1}^t \sum_{n=1}^N \mathbf{x}_{n,j}$, $\boldsymbol{\Lambda}_0 = 0.02 \mathbf{I}_4$, for all $n = 1, \ldots, N$. We set $N = 20$ as number of chains, $L = 20$ as number of tries, and $E_T \approx 8700$ as total number of evaluations of the target. For a fair comparison, we again consider $N$ and $T$ for IPCs such that $E'_T = NT$ is equal to $E_T$, i.e., $E'_T = E_T$. The vertical proposal pdfs are the same than those for the IPCs scheme. Furthermore, we apply a data-tempering approach [54] described in Section 6, employing a sequence of 29 target pdfs $\pi_i$ each one considering an increasing number of observations $K_i = 2 + (i - 1)$ with $i = 1, \ldots, 29$. The computational effort $E_T$ is distributed uniformly in each $\pi_i$. We compute the Relative Error (RE) of the last states of the $N$ chains with respect to the true vector $\mathbf{f}$. Figure 8 depicts the curves of the RE versus different values of $\sigma \in [0.05, 0, 5]$. We can observe that O-MCMC-PMTM always outperforms IPCs in this optimization problem.

---

[8] Let us remark that the estimation of all these parameters would make the inference harder, but can be easily incorporated into our algorithm.

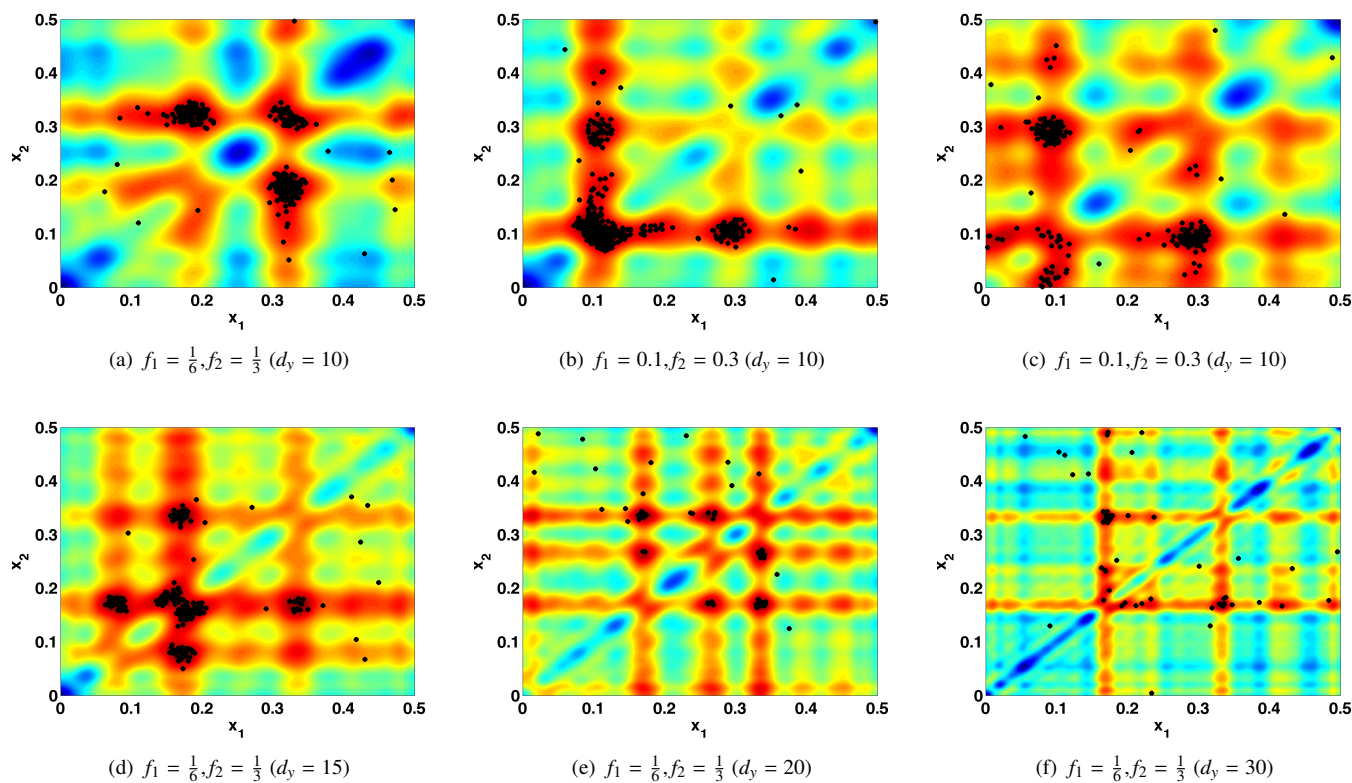

(a) $f_1 = \frac{1}{6}, f_2 = \frac{1}{3}$ ($d_y = 10$)   (b) $f_1 = 0.1, f_2 = 0.3$ ($d_y = 10$)   (c) $f_1 = 0.1, f_2 = 0.3$ ($d_y = 10$)

(d) $f_1 = \frac{1}{6}, f_2 = \frac{1}{3}$ ($d_y = 15$)   (e) $f_1 = \frac{1}{6}, f_2 = \frac{1}{3}$ ($d_y = 20$)   (f) $f_1 = \frac{1}{6}, f_2 = \frac{1}{3}$ ($d_y = 30$)

Figure 6. Several examples of function $V(\mathbf{x}) = \log \pi(\mathbf{x})$ with $d_x = 2$, given different realizations of the measurements $y[1], \ldots, y[K]$. In Figures (a)-(d)-(e)-(f), we set $\mathbf{f} = [\frac{1}{6}, \frac{1}{3}]^\top$ and consider $d_y = 10, 15, 20, 30$ observations, respectively. In Figures (b)-(c), we set $\mathbf{f} = [0.1, 0.3]^\top$ and $d_y = 10$ observations. Black dotted points shows all the states generated throughout an O-MCMC-PMTM run with $N = 10$, $L = 10$ and $T = 500$. The initial states are chosen uniformly within $\mathcal{D} = [0, \frac{1}{2}]^\top$.

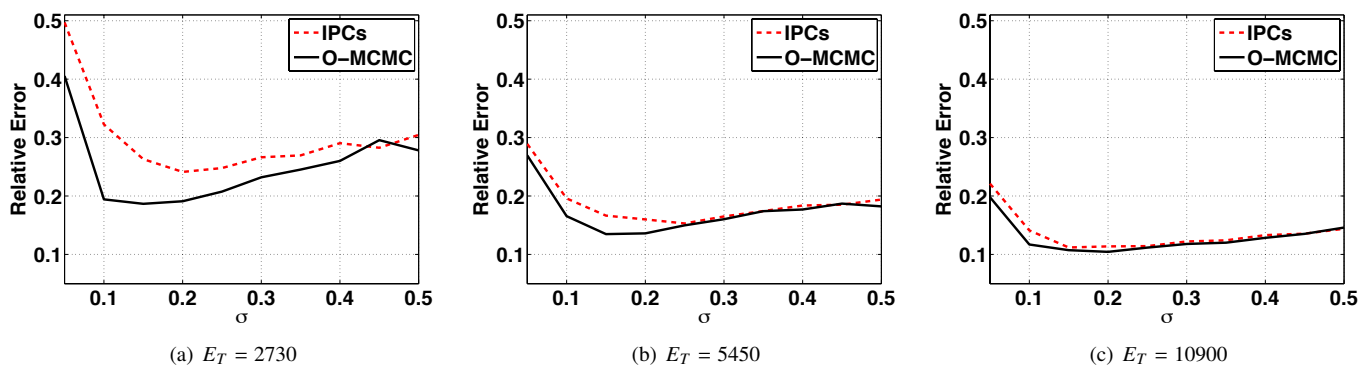

(a) $E_T = 2730$                (b) $E_T = 5450$                (c) $E_T = 10900$

Figure 7. Relative Error (averaged over 500 runs) in the first experiment for O-MCMC-SMH (solid line) and IPCs (dashed line) with different computational effort $E_T$. Note that O-MCMC always outperforms IPCs. Note also that their performance becomes similar as the overall computational cost $E_T$ grows (due to the small size of the state space, $\mathcal{D} = \left[0, \frac{1}{2}\right]^2$).

**Third experiment.** Now we test O-MCMC in many different dimensions $d_x \in \{2, 3, \ldots, 30\}$. The ground truth is the $d_x$-dimensional vector $\mathbf{f} = [f_1, f_2, \ldots, f_{d_x}]^\top$, with $f_i = \frac{i}{2(d_x+1)}$, i.e., equally spaced within the interval $[0\ 0.5]$. The problem is again finding the global mode of the target $\pi$, which in this case includes $d_y = 200$ observations. We consider $\mathbf{f}$ as true localization of the global mode for similar aforementioned reasons. The search is now restricted to the subset of $\mathbb{R}^{d_x}$ where the dimensions of $\mathbf{x}$ are decreasingly sorted. Now the proposed O-MCMC-PMTM uses Gaussian horizontal proposals in the mixture $\psi$, with $\Lambda = \lambda^2 \mathbf{I}_{d_x}$ for all $n$. Suggested by the second experiment, we set $\lambda = 0.5$ for the horizontal steps, and $\sigma = 0.25$ for the vertical chains. We set $N = 50$ as number of chains, $L = 10$ as number of tries, and $E_T \approx 47040$ as total number of evaluations of the target. For a fair comparison, we again consider $N$ and $T$ for IPCs such that $E'_T = NT$ is equal to $E_T$, i.e., $E'_T = E_T$. The vertical proposal pdfs are the same than those for the IPCs scheme. The data-tempering approach of [54] is implemented with a sequence of 7 target pdfs $\pi_i$ each one considering an increasing number of observations $\left[d_y^{(1)}, ..., d_y^{(7)}\right] = [2, 5, 10, 20, 50, 100, 200]$ with $i = 1, \ldots, 7$. We compute the MSE (adding the MSE of every dimension) and the Relative Error (RE) of the last states of the $N = 50$ chains with respect to the true vector $\mathbf{f}$. Figure 9 depicts the curves of the MSE and RE versus different values of $\sigma \in [0.05, 0, 5]$. Note that the performance degrades when the dimension is increased, but O-MCMC again always outperforms IPCs.

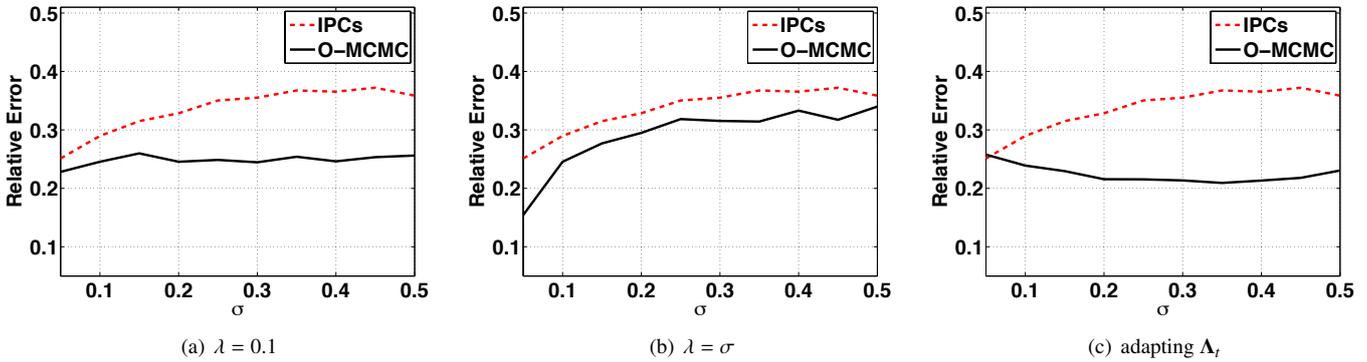

Figure 8. Relative Error (averaged over 500 runs) in the second experiment (dimension $d_x = 4$) searching the global maximum with O-MCMC (solid line) and IPCs (dashed line). In the O-MCMC-PMTM scheme, we use different $\lambda = 0.1$ and $\lambda = \sigma$, for the covariance matrices $\Lambda_n = \lambda^2 \mathbf{I}_4$ of the horizontal proposal pdfs in the mixture $\psi$. Moreover, we test an adapted covariance matrices for the horizontal proposal pdfs (see Figure (c)).

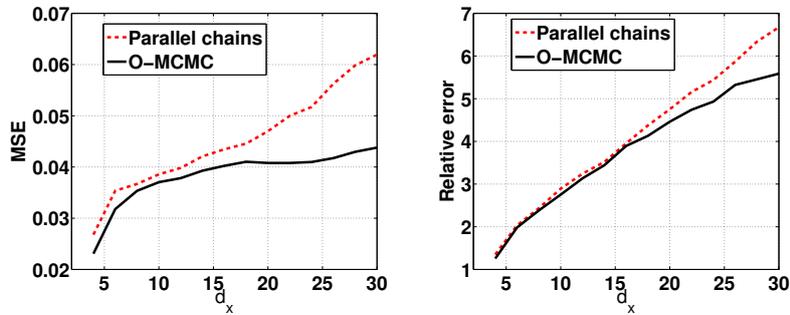

Figure 9. MSE and relative Error (averaged over 100 runs) as function of the dimension $d_x$, in the third experiment (with $d_y = 200$ observations) for O-MCMC-SMH (solid line) and IPCs (dashed line) with the same computational effort $E_T$. Note that the performance degrades when the dimension is increased, but O-MCMC always outperforms IPCs.

### 8.3. Localization in a Wireless Sensor Network

In this section, we address the problem of positioning a static target in the two-dimensional space of a wireless sensor network using only range measurements. More specifically, we consider a random vector $\mathbf{X} = [X_1, X_2]^\top$ to denote the target's position in the $\mathbb{R}^2$ plane. The position is then a specific realization $\mathbf{x}$. The measurements are obtained from 6 range sensors located at $\mathbf{h}_1 = [1, -8]^\top$, $\mathbf{h}_2 = [8, 10]^\top$, $\mathbf{h}_3 = [-15, -7]^\top$, $\mathbf{h}_4 = [-8, 1]^\top$, $\mathbf{h}_5 = [10, 0]^\top$ and $\mathbf{h}_6 = [0, 10]^\top$. The measurement equations are

$$Y_{j,r} = -20 \log\left(\|\mathbf{x} - \mathbf{h}_j\|^2\right) + \Theta_j, \quad j = 1, \ldots, 6, \quad r = 1, \ldots, d_y, \tag{17}$$

where $\Theta_j \sim \mathcal{N}(\theta_j | \mathbf{0}, \omega_j^2 \mathbf{I})$, with $\omega_j = 5$ for all $j \in 1, \ldots, 6$. Note that the total number of data is $6d_y$. We consider a vague Gaussian prior pdf with mean $[0, 0]^\top$ and covariance matrix $[\omega_0^2 \, 0; 0 \, \omega_0^2]^\top$ with $\omega_0 = 10$,

We simulate $6d_y = 360$ measurements from the model ($d_y = 60$ observations from each sensor), fixing $x_1 = 3.5$ and $x_2 = 3.5$. Our goal is to compute the expected value of the posterior $\bar{\pi}(\mathbf{x}|\mathbf{y})$, using different Monte Carlo techniques. Since we consider a fixed sequence of observations, for comparing the performance of the different methods, we first approximate the expected value of $\bar{\pi}(\mathbf{x}|\mathbf{y})$ using an extremely thin grid obtaining $E[\mathbf{X}] \approx [3.415, 3.539]^\top$ (so that we compare the Monte Carlo approximation with these true values).

We compare O-MCMC-SMH with SMC in App. B.2 both with adaptation covariance matrix of the proposal in the "horizontal" step (i.e., used in the resampling in SMC), as suggested in Section 5.5. In both cases, we consider MH vertical kernels, with $q_n(\mathbf{x}|\mathbf{x}_{n,t-1}) = \mathcal{N}(\mathbf{x}; \mathbf{x}_{n,t-1}, \mathbf{C}_n)$ as proposal pdfs, using the same isotropic covariance matrix, $\mathbf{C}_n = \sigma^2 \mathbf{I}_2$, for all $n = 1, \ldots, N$, and $\sigma \in \{1, 2\}$. As horizontal proposal in O-MCMC-SMH we use a Gaussian pdf, $\varphi_t(\mathbf{x}) = \mathcal{N}(\mathbf{x}; \boldsymbol{\mu}_t, \boldsymbol{\Lambda}_t)$ where $\boldsymbol{\mu}_t$ and $\boldsymbol{\Lambda}_t$ are adapted online, $\boldsymbol{\mu}_t = \frac{1}{Nt} \sum_{j=1}^t \sum_{n=1}^N \mathbf{x}_{n,j}$, and $\boldsymbol{\Lambda}_t = \frac{1}{Nt} \sum_{j=1}^t \sum_{n=1}^N (\mathbf{x}_{n,j} - \boldsymbol{\mu}_t)(\mathbf{x}_{n,j} - \boldsymbol{\mu}_t)^\top + \boldsymbol{\Lambda}_0$, where $\boldsymbol{\mu}_0 = [0, 0]^\top$, $\boldsymbol{\Lambda}_0 = \lambda^2 \mathbf{I}_2$ with $\lambda = 0.2$, and $T_{train} = T_V$. The "horizontal" proposals in SMC are $\varphi_n(\mathbf{x}|\mathbf{x}_{n,t}, \boldsymbol{\Lambda}_t) = \mathcal{N}(\mathbf{x}; \mathbf{x}_{n,t}, \boldsymbol{\Lambda}_t)$, for $n = 1, \ldots, N$, and $\boldsymbol{\Lambda}_t = \frac{1}{Nt} \sum_{j=1}^t \sum_{n=1}^N (\mathbf{x}_{n,j} - \boldsymbol{\mu}_t)(\mathbf{x}_{n,j} - \boldsymbol{\mu}_t)^\top + \boldsymbol{\Lambda}_0$ is adapted as in O-MCMC-SMH.

We set $T_H = 1$ and number of epochs $M = 100$, for both algorithms. Then, we test different values of $N$ parallel chains, and vertical steps $T_V$. The total number of evolution of the target is $E_T = M(NT_V + T_H) = 100(NT_V + 1)$ for O-MCMC-SMH and $E_T = 100N(T_V + 1)$ for SMC (see Appendix B.2). We repeat the experiments 200 times (with independent runs) and average the results. At each run, the initial states are chosen randomly, $\mathbf{x}_{n,0} \sim \mathcal{U}([-10, 10] \times [-10, 10])$ for $n = 1, \ldots, N$. Table 12 gives the MSE in estimation of the expected value of the posterior, with the different methods. The results in Table 12 confirm that O-MCMC outperforms SMC even with less computational cost (a smaller $E_T$). This shows that the advantage of replacing the resampling procedure with an orthogonal MCMC technique like SMH, in this case.

| | O-MCMC with SMH | | | | | | adaptive SMC | | | | | |
|---|---|---|---|---|---|---|---|---|---|---|---|---|
| $N$ | 2 | | 5 | | 10 | | 5 | | 10 | | 100 | |
| $T_V$ | 2 | 20 | 2 | 20 | 2 | 20 | 2 | 20 | 2 | 20 | 2 | 20 |
| $\sigma = 1$ | 0.1507 | 0.0021 | 0.0923 | 0.0008 | 0.0522 | 0.0003 | 9.2133 | 0.0233 | 9.8118 | 0.0065 | 4.0167 | 0.0022 |
| $\sigma = 2$ | 0.1546 | 0.0029 | 0.0951 | 0.0011 | 0.0536 | 0.0004 | 3.8216 | 0.0684 | 3.6501 | 0.0593 | 1.1419 | 0.0370 |
| $E_T$ | 500 | 4100 | 1100 | 10100 | 2100 | 20100 | 1500 | 10500 | 3000 | 21000 | 30000 | $21 \cdot 10^4$ |

Table 12. Mean Square Error (MSE) in the estimation of the target position, using O-MCMC-SMH ($T_H = 1$) and SMC. The total number of evaluations of the target is $E_T = 100(NT_V + 1)$ for O-MCMC-SMH, whereas $E_T = 100N(T_V + 1)$ for adaptive SMC.

## 9. Conclusions

In this work, we have introduced a novel family of MCMC algorithms, named Orthogonal MCMC schemes, that incorporates "horizontal" MCMC transitions to share information among a cloud of parallel "vertical" MCMC chains. We have described different alternatives for exchanging information among independent parallel chains. Compared to the fully independent parallel chains approach, the novel interacting techniques show a more robust behavior with respect to the parameterization and better performance for different number of chains. One reason of this behavior is that the novel algorithms provide a good trade-off between the use of an independent and a random walk proposal density, i.e., between local an global exploration. We have considered two different approaches for the interaction among the chains: in the first one, an MCMC technique over the entire population is directly applied, whereas in the second one, the initial population $\mathcal{P}_t$ is used for building a suitable mixture density $\psi(\mathbf{x})$ employed as proposal function in the horizontal transitions. This second approach can be interpreted as an adaptive MCMC scheme, where the location parameters of the $N$ components of the mixture $\psi(\mathbf{x})$ are driven by $N$ parallel MCMC chains. The outputs of these parallel chains are also employed in the approximation of the target. Furthermore, we have designed different parallel Multiple Try Metropolis (P-MTM) schemes using an independent proposal pdf, where the generated candidates are recycled in order to reduce the overall computational cost. Finally, we have described two modified versions of O-MCMC for optimization and inference in big data problems. The ergodicity of all the proposed methodologies has been proved and several numerical simulations have been provided in order to show the advantages of the novel approach.

In future works we plan to address the development of parallel and data-distributed implementations of O-MCMC algorithms, considering the use of strategies for monitoring the convergence of the vertical chains, and tempered versions for sampling from high-dimensional and multi-modal targets.

## 10. ACKNOWLEDGEMENTS

This work has been supported by the ERC grant 239784 and AoF grant 251170, the Spanish government through the DISSECT project, the BBVA Foundation through the MG-FIAR project, by the Grant 2014/23160-6 of São Paulo Research Foundation (FAPESP) and by the Grant 305361/2013-3 of National Council for Scientific and Technological Development (CNPq).

## A. Stationary distribution of O-MCMC

In this section, we prove the ergodicity of the proposed schemes. First of all, we study the mixture-based approach introduced in Section 5.2, and then the population-based approach described in Section 5.1, within O-MCMC.

### A.1. Analysis for the mixture-based approach

Let us consider two MCMC kernels, $K_n^{(V)}(\mathbf{y}|\mathbf{x})$ and $K_n^{(H)}(\mathbf{z}|\mathbf{y})$ with $\mathbf{x}, \mathbf{y}, \mathbf{z} \in \mathcal{D} \in \mathbb{R}^{d_x}$, corresponding to the $n$-th chain for the vertical and horizontal steps, respectively. We assume $\bar{\pi}(\cdot)$ is the invariant density of both chains. Namely, we consider MCMC techniques whose steps are summarized in the two conditional probabilities, $K_n^{(V)}(\mathbf{y}|\mathbf{x})$ and $K_n^{(H)}(\mathbf{z}|\mathbf{y})$, such that

$$\int_{\mathcal{D}} K_n^{(V)}(\mathbf{y}|\mathbf{x})\bar{\pi}(\mathbf{x})d\mathbf{x} = \bar{\pi}(\mathbf{y}),$$

$$\int_{\mathcal{D}} K_n^{(H)}(\mathbf{z}|\mathbf{y})\bar{\pi}(\mathbf{y})d\mathbf{y} = \bar{\pi}(\mathbf{z}).$$

For the sake of simplicity, we tackle a simpler case where $K_n^{(V)}$, $K_n^{(H)}$ are used sequentially, once each (i.e., $T_V = 1$ and $T_H = 1$). Namely, we consider the sequential application of $K_n^{(V)}$ and $K_n^{(H)}$, i.e, first draw $\mathbf{y}' \sim K_n^{(V)}(\mathbf{y}|\mathbf{x})$ and then draw $\mathbf{z}' \sim K_n^{(H)}(\mathbf{z}|\mathbf{y}')$. The transition probability from $\mathbf{z}$ to $\mathbf{x}$ is given by

$$T(\mathbf{z}|\mathbf{x}) = \int_{\mathcal{D}} K_n^{(H)}(\mathbf{z}|\mathbf{y}) K_n^{(V)}(\mathbf{y}|\mathbf{x})d\mathbf{y}. \tag{18}$$

The target $\bar{\pi}$ is also invariant w.r.t. $T(\mathbf{z}|\mathbf{x})$ [38, Chapter 1]. Indeed, we can write

$$
\begin{aligned}
\int_{\mathcal{D}} T(\mathbf{z}|\mathbf{x})\bar{\pi}(\mathbf{x})d\mathbf{x} &= \\
&= \int_{\mathcal{D}} \left[\int_{\mathcal{D}} K_n^{(H)}(\mathbf{z}|\mathbf{y}) K_n^{(V)}(\mathbf{y}|\mathbf{x})d\mathbf{y}\right] \bar{\pi}(\mathbf{x})d\mathbf{x}, \\
&= \int_{\mathcal{D}} K_n^{(H)}(\mathbf{z}|\mathbf{y}) \left[\int_{\mathcal{D}} K_n^{(V)}(\mathbf{y}|\mathbf{x})\bar{\pi}(\mathbf{x})d\mathbf{x}\right] d\mathbf{y}, \\
&= \int_{\mathcal{D}} K_n^{(H)}(\mathbf{z}|\mathbf{y})\bar{\pi}(\mathbf{y})d\mathbf{y}, \\
&= \bar{\pi}(\mathbf{z}),
\end{aligned} \tag{19}
$$

which is precisely the definition of invariant pdf associated to $T(\mathbf{z}|\mathbf{x})$. Clearly, this argument is valid for each $n = 1 \ldots, N$, and can be easily extended for the product of more than two kernels (i.e., for any $T_V, T_H < \infty$).

### A.2. Analysis for the population-based approach

Considering now an extended state space, $\mathbb{R}^{d_x \times N}$, we can interpret that O-MCMC yields a unique chain in $\mathbb{R}^{d_x \times N}$. Namely, one population of states at the $t$-th iteration represents one extended state of this unique chain. Here, we show that this chain, generated by O-MCMC, has the extended target density

$$\bar{\pi}_g(\mathbf{x}_1, \ldots, \mathbf{x}_N) \propto \prod_{n=1}^{N} \pi(\mathbf{x}_n),$$

as invariant pdf. We can use similar arguments to those employed previously, considering now a population of current states, i.e.,

$$\mathcal{P}_{t-1} = \{\mathbf{x}_{1,t-1}, \ldots, \mathbf{x}_{N,t-1}\}.$$

We denote the vertical MCMC kernels as $K_n^{(V)}(\mathbf{x}_{n,t}|\mathbf{x}_{n,t-1})$ with $\bar{\pi}$ as invariant pdf, whereas $K^{(H)}(\mathcal{P}_t|\mathcal{P}_{t-1})$ denotes the horizontal kernel[9] with invariant pdf the aforementioned extended target pdf,

$$\bar{\pi}_g(\mathcal{P}) \propto \pi_g(\mathbf{x}_1, \ldots, \mathbf{x}_N) = \prod_{n=1}^{N} \pi(\mathbf{x}_n).$$

Thus, in this case the complete kernel of the O-MCMC procedure formed by one vertical and one orthogonal step, is

$$T(\mathcal{P}_t|\mathcal{P}_{t-2}) \quad = \quad \int_{\mathcal{D}^N} K^{(H)}(\mathcal{P}_t|\mathcal{P}_{t-1}) \left[ \prod_{n=1}^{N} K_n^{(V)}(\mathbf{x}_{n,t-1}|\mathbf{x}_{n,t-2}) \right] \prod_{n=1}^{N} d\mathbf{x}_{n,t-1}.$$

In this case, we can write

$$\int_{\mathcal{D}^N} T(\mathcal{P}_t|\mathcal{P}_{t-2}) \bar{\pi}_g(\mathcal{P}_{t-2}) d\mathcal{P}_{t-2} =$$

$$= \int_{\mathcal{D}^N} K^{(H)}(\mathcal{P}_t|\mathcal{P}_{t-1}) \left[ \int_{\mathcal{D}^N} \left( \prod_{n=1}^{N} K_n^{(V)}(\mathbf{x}_{n,t-1}|\mathbf{x}_{n,t-2}) \prod_{n=1}^{N} \bar{\pi}(\mathbf{x}_{n,t-2}) \right) \prod_{n=1}^{N} d\mathbf{x}_{n,t-2} \right] \prod_{n=1}^{N} d\mathbf{x}_{n,t-1},$$

$$= \int_{\mathcal{D}^N} K^{(H)}(\mathcal{P}_t|\mathcal{P}_{t-1}) \prod_{n=1}^{N} \bar{\pi}(\mathbf{x}_{n,t-1}) \prod_{n=1}^{N} d\mathbf{x}_{n,t-1}$$

$$= \int_{\mathcal{D}^N} K^{(H)}(\mathcal{P}_t|\mathcal{P}_{t-1}) \bar{\pi}_g(\mathcal{P}_{t-1}) d\mathcal{P}_{t-1} = \bar{\pi}_g(\mathcal{P}_t).$$

Namely, the kernel $T(\mathcal{P}_t|\mathcal{P}_{t-2})$ has $\bar{\pi}_g$ as invariant density. Once more, this result can be easily extended when $T_V$ vertical and $T_H$ horizontal transitions are applied by using the same arguments. Note that the generated parallel chains preserve the pdf $\pi$ as invariant pdf, as shown previously, but in general is not reversible [38, Section 1.12.7].

### B. Population Monte Carlo (PMC), Sequential Monte Carlo (SMC) and distribution after resampling

Resampling procedures are employed in different Monte Carlo techniques such as Population Monte Carlo (PMC), Iterated Batch Importance Sampler (IBIS) and, more generally, in Sequential Monte Carlo (SMC) methods for a static scenario [13, 54, 15].

#### B.1. Standard PMC

For simplicity, let us consider here a standard PMC-type scheme. In PMC, $N$ different proposal pdfs $q_1, \ldots, q_N$ are employed at each iteration. Starting from $\{\mathbf{x}_{1,0}, \ldots, \mathbf{x}_{N,0}\}$, the basic PMC scheme consists of the following steps:

1. For $t = 1, \ldots, T$:
   (a) For $n = 1, \ldots, N$:
      i. **Propagation:** Draw one sample $\mathbf{x}_{n,t}$ from $q_n$, i.e.,

      $$\mathbf{x}_{n,t} \sim q_n(\mathbf{x}|\tilde{\mathbf{x}}_{n,t-1}),$$

      ii. **Weighting:** Assign the unnormalized weight $w_{n,t} = \frac{\pi(\mathbf{x}_{n,t})}{q_n(\mathbf{x}_{n,t}|\tilde{\mathbf{x}}_{n,t-1})}$ and store the pair $\{\mathbf{x}_{n,t}, w_{n,t}\}$.
      iii. **Resampling:** Draw $N$ independent samples $\{\mathbf{z}_1, \ldots, \mathbf{z}_N\}$ such that each $\mathbf{z}_n \in \{\mathbf{x}_{1,t}, \ldots, \mathbf{x}_{N,t}\}$ for $n = 1, \ldots, N$, with probability

      $$\beta_n = \frac{w_{n,t}}{\sum_{k=1}^{N} w_{k,t}} = \frac{\frac{\pi(\mathbf{x}_{n,t})}{q_n(\mathbf{x}_{n,t}|\tilde{\mathbf{x}}_{n,t-1})}}{\sum_{k=1}^{N} \frac{\pi(\mathbf{x}_{k,t})}{q_k(\mathbf{x}_{k,t}|\tilde{\mathbf{x}}_{k,t-1})}}. \tag{20}$$

      iv. Set $\tilde{\mathbf{x}}_{n,t} = \mathbf{z}_n$.
2. Use all the pairs $\{\mathbf{x}_{n,t}, w_{n,t}\}_{n,t=1}^{N,T}$ in order to build a unique IS estimator (normalizing jointly the weights $w_{n,t}$).

The step 2(b) corresponds to resample (with replacement) $N$ times the population $\{\mathbf{x}_{n,t}\}_{n=1}^{N}$. Note that the weights in Eq. (20) are the same used in (10).

---

[9]For the sake of simplicity, we abuse of the notation using here the set $\mathcal{P}_t$ as a vector. Moreover, we assume $T_V = 1$ and $T_H = 1$.

### B.2. A Sequential Monte Carlo sampler

A generic Sequential Monte Carlo method for a static inference scenario have been exhaustively described in [15]. For facilitating the comparison with O-MCMC, we describe a specific SMC scheme (also known as "resample-move" scheme [62]) without considering a sequence of tempered target pdfs, but always the true target, i.e, $\bar{\pi}$. Below, we describe a specific SMC technique which belongs to this wide class, due to its connection to the O-MCMC framework. Starting from the population $\{\mathbf{x}_{1,0}, \dots, \mathbf{x}_{N,0}\}$, this specific SMC scheme for a static scenario consists of the following steps:

1. For $t = 1, \dots, T$:
   (a) For $n = 1, \dots, N$:
      i. **Propagation:** draw one sample $\mathbf{x}_{n,t}$ from $q_n$, i.e.,

      $$\mathbf{x}_{n,t} \sim q_n(\mathbf{x}|\tilde{\mathbf{x}}_{n,t-1}),$$

      ii. **Weighting:** Assign the unnormalized weight $w_{n,t} = \frac{\pi(\mathbf{x}_{n,t})}{q_n(\mathbf{x}_{n,t}|\tilde{\mathbf{x}}_{n,t-1})}$ and store the pair $\{\mathbf{x}_{n,t}, w_{n,t}\}$.
      iii. **Resampling:** Draw $N$ independent samples $\{\mathbf{z}_1, \dots, \mathbf{z}_N\}$ such that each $\mathbf{z}_n \in \{\mathbf{x}_{1,t}, \dots, \mathbf{x}_{N,t}\}$ for $n = 1, \dots, N$, with probability

      $$\beta_n = \frac{w_{n,t}}{\sum_{k=1}^{N} w_{k,t}} = \frac{\frac{\pi(\mathbf{x}_{n,t})}{q_n(\mathbf{x}_{n,t}|\tilde{\mathbf{x}}_{n,t-1})}}{\sum_{k=1}^{N} \frac{\pi(\mathbf{x}_{k,t})}{q_k(\mathbf{x}_{k,t}|\tilde{\mathbf{x}}_{k,t-1})}}. \tag{21}$$

      iv. **Moving:** Apply one step of $N$ parallel MCMC chains (with invariant target $\bar{\pi}$), starting from $\{\mathbf{z}_1, \dots, \mathbf{z}_N\}$ and obtaining the new population $\{\tilde{\mathbf{x}}_{1,t}, \dots, \tilde{\mathbf{x}}_{N,t}\}$.

Above we have considered only $T_V = 1$ step for each "vertical" MCMC moves. However, it can be used $T_V$ different steps as well, as in the numerical example in Section 8.3. In this case, the total number of evaluations of the target is $E_T = TN(1 + T_V)$. Finally, observe that the resampling plays a similar role to the orthogonal step in O-MCMC.

### B.3. Distribution after resampling

For the sake of simplicity, since we consider a generic iteration $t$, let us simplify the notation, denoting $\mathbf{x}_n = \mathbf{x}_{n,t} \sim q_n(\mathbf{x}|\mathbf{x}_{n,t-1})$ $(1 \le n \le N, 1 \le t \le T)$, and $q_n(\mathbf{x}) = q_n(\mathbf{x}|\mathbf{x}_{n,t-1})$. Moreover, we consider the following simplified procedure:

1. For $n = 1, \dots, N$, draw $\mathbf{x}_n \sim q_n(\mathbf{x})$ and compute the weights $\beta_n$ in Eq. (20).
2. Resample one sample $\mathbf{z} \in \{\mathbf{x}_1, \dots, \mathbf{x}_N\}$ according to the probabilities $\beta_n$, $n = 1, \dots, N$.

In this section, we write the density $\phi(\mathbf{z})$ where $\mathbf{z}$ is obtained by the procedure above. We define as

$$\mathbf{m}_{\neg n} = [\mathbf{x}_1, \dots, \mathbf{x}_{n-1}, \mathbf{x}_{n+1}, \dots, \mathbf{x}_N],$$

the matrix containing all the samples except for the $n$-th. Let us also denote as $\mathbf{z} \in \{\mathbf{x}_1 \dots, \mathbf{x}_N\}$, a generic sample after applying one multinomial resampling step. Hence, the distribution of $\mathbf{z}$ is given by

$$\phi(\mathbf{z}) = \int_{\mathcal{D}^N} \hat{\pi}^{(N)}(\mathbf{z}|\mathbf{x}_1, \dots, \mathbf{x}_N) \left[ \prod_{n=1}^{N} q_n(\mathbf{x}_n) \right] d\mathbf{x}_1 \dots d\mathbf{x}_N, \tag{22}$$

where

$$\hat{\pi}^{(N)}(\mathbf{z}|\mathbf{x}_1, \dots, \mathbf{x}_N) = \sum_{j=1}^{N} \beta_j \delta(\mathbf{z} - \mathbf{x}_j), \tag{23}$$

and $\beta_j$ are given in Eq. (20). Note that, by using the notation $\hat{\pi}^{(N)}(\mathbf{z}|\mathbf{x}_1, \dots, \mathbf{x}_N)$ we have emphasized the dependence on the generated samples $\mathbf{x}_n$'s in order to facilitate the understanding of Eq. (22). After some straightforward rearrangements, Eq. (22) can be rewritten as

$$\phi(\mathbf{z}) = \sum_{j=1}^{N} \left( \int_{\mathcal{D}^{N-1}} \frac{\pi(\mathbf{z})}{\sum_{n=1}^{N} \frac{\pi(\mathbf{x}_n)}{q_n(\mathbf{x}_n)}} \left[ \prod_{\substack{n=1 \\ n \neq j}}^{N} q_n(\mathbf{x}_n) \right] d\mathbf{m}_{\neg j} \right). \tag{24}$$

Finally, we can write

$$\phi(\mathbf{z}) = \pi(\mathbf{z}) \sum_{j=1}^{N} \int_{\mathcal{D}^{N-1}} \frac{1}{N\hat{Z}} \left[ \prod_{\substack{n=1 \\ n \neq j}}^{N} q_n(\mathbf{x}_n) \right] d\mathbf{m}_{\neg j}, \tag{25}$$

where $\hat{Z} = \frac{1}{N} \sum_{n=1}^{N} \frac{\pi(\mathbf{x}_n)}{q_n(\mathbf{x}_n)}$ is the estimate of the normalizing constant of the target obtained by using the importance sampling technique. The equation above represents the density of a resampled particle. Clearly, for a finite value of $N$, there exists a discrepancy between $\phi(\mathbf{z})$ and $\tilde{\pi}(\mathbf{z})$.

### C. Calderhead's MCMC technique based on multiple candidates

A similar approach to ensemble algorithm proposed in Table 5 has been proposed in [12], which is suggested as a general construction in order to parallelize a Metropolis-Hastings algorithm. The algorithm can be employed as orthogonal technique in O-MCMC and is outlined below:

1. Set $t = 1$, and choose an initial state $\mathbf{x}_1$.
2. For $m = 1, \ldots, \frac{T}{L}$:
   (a) Draw $L$ candidates $\mathbf{z}_1, \ldots, \mathbf{z}_L$ from $q(\mathbf{z}_1, \ldots, \mathbf{z}_L | \mathbf{x}_t)$ and set $\mathbf{z}_{L+1} = \mathbf{x}_t$.
   (b) Denoting with

   $$\mathbf{Z}_{\neg k} = [\mathbf{z}_1, \ldots, \mathbf{z}_{k-1}, \mathbf{z}_{k+1}, \ldots, \mathbf{z}_L],$$

   the matrix containing as column all the $\mathbf{z}$'s with the exception of $\mathbf{z}_k$, Compute $L + 1$ normalized weights

   $$\beta_k = \frac{\pi(\mathbf{z}_k) q(\mathbf{Z}_{\neg k} | \mathbf{z}_k)}{\sum_{\ell=1}^{L+1} \pi(\mathbf{z}_\ell) q(\mathbf{Z}_{\neg \ell} | \mathbf{z}_\ell)}, \tag{26}$$

   (c) Resample $L$ times within the set $\{\mathbf{z}_1, \ldots, \mathbf{z}_{L+1}\}$ of $L + 1$ elements, obtaining $L$ samples $\mathbf{x}_{t+1}, \ldots, \mathbf{x}_{t+L}$, according to the probabilities $\beta_\ell$ in Eq. (26), $\ell = 1, \ldots, L + 1$.
   (d) Set $t = t + L = mL$.
3. Return $\{\mathbf{x}_t\}_{t=1}^{T}$.

In this case, unlike in Table 5, the proposal density $q$ depends on the previous state of the chain. Below, we also discuss the ergodicity of this technique where, for simplicity, we consider of resampling only once in step 2(c) and assume

$$q(\mathbf{z}_1, \ldots, \mathbf{z}_L | \mathbf{x}_t) = \prod_{\ell=1}^{L} q(\mathbf{z}_\ell | \mathbf{x}_t).$$

Denoting as $K(\mathbf{z}_k | \mathbf{x}_t)$ the kernel of the method and noting that $q(\mathbf{z}_1, \ldots, \mathbf{z}_L | \mathbf{x}_t) = q(\mathbf{Z}_{\neg(L+1)} | \mathbf{x}_t)$, we can write

$$
\begin{aligned}
\tilde{\pi}(\mathbf{x}_t) K(\mathbf{z}_k | \mathbf{x}_t) &= L \tilde{\pi}(\mathbf{x}_t) \int_{\mathcal{D}^{L-1}} q(\mathbf{z}_1, \ldots, \mathbf{z}_L | \mathbf{x}_t) \frac{\pi(\mathbf{z}_k) q(\mathbf{Z}_{\neg k} | \mathbf{z}_k)}{\sum_{\ell=1}^{L+1} \pi(\mathbf{z}_\ell) q(\mathbf{Z}_{\neg \ell} | \mathbf{z}_\ell)} d\mathbf{Z}_{\neg k}, \\
&= \frac{L}{Z} \pi(\mathbf{x}_t) \pi(\mathbf{z}_k) \int_{\mathcal{D}^{L-1}} \frac{q(\mathbf{Z}_{\neg(L+1)} | \mathbf{z}_t) q(\mathbf{Z}_{\neg k} | \mathbf{z}_k)}{\sum_{\ell=1}^{L+1} \pi(\mathbf{z}_\ell) q(\mathbf{Z}_{\neg \ell} | \mathbf{z}_\ell)} d\mathbf{Z}_{\neg k} =
\end{aligned}
$$

Since $\mathbf{z}_{L+1} = \mathbf{x}_t$, then

$$
\begin{aligned}
\tilde{\pi}(\mathbf{x}_t) K(\mathbf{z}_k | \mathbf{x}_t) &= \frac{L}{Z} \pi(\mathbf{x}_t) \pi(\mathbf{z}_k) \int_{\mathcal{D}^{L-1}} \frac{q(\mathbf{Z}_{\neg(L+1)} | \mathbf{x}_t) q(\mathbf{Z}_{\neg k} | \mathbf{z}_k)}{\sum_{\ell \neq k, L+1}^{L} \pi(\mathbf{z}_\ell) q(\mathbf{Z}_{\neg \ell} | \mathbf{z}_\ell) + \pi(\mathbf{z}_k) q(\mathbf{Z}_{\neg k} | \mathbf{z}_k) + \pi(\mathbf{x}_t) q(\mathbf{Z}_{\neg(L+1)} | \mathbf{x}_t)} d\mathbf{Z}_{\neg k}, \\
&= \tilde{\pi}(\mathbf{z}_k) K(\mathbf{x}_t | \mathbf{z}_k),
\end{aligned}
$$

which is the detailed balance condition (we have used that we can exchange the position of $\mathbf{z}_k$ and $\mathbf{x}_t$ without varing the expression above).

## D. Ergodicity of SMH

Let us recall that we denote as $\mathcal{P}_{t-1} = \{\mathbf{x}_{1,t-1}, \ldots, \mathbf{x}_{N,t-1}\}$, the population of the states at the $(t-1)$-th iteration. A sufficient condition for proving the ergodicity of the chain, generated by SMH, is given by the detailed balance condition with respect to the extended target $\bar{\pi}_g(\mathbf{x}_1, \ldots, \mathbf{x}_N) = \prod_{i=1}^{N} \bar{\pi}_i(\mathbf{x}_i)$. For the case $\mathcal{P}_t \neq \mathcal{P}_{t-1}$ (the case $\mathcal{P}_t = \mathcal{P}_{t-1}$ is straightforward), the kernel of SMH can be expressed as

$$K(\mathcal{P}_t|\mathcal{P}_{t-1}) = N\varphi(\mathbf{x}_{0,t-1}) \frac{\frac{\varphi(\mathbf{x}_{j,t-1})}{\pi(\mathbf{x}_{j,t-1})}}{\sum_{i=1}^{N} \frac{\varphi(\mathbf{x}_{i,t-1})}{\pi(\mathbf{x}_{i,t-1})}} \alpha(\mathcal{P}_{t-1}, \mathbf{x}_{0,t-1}),$$

where we have considered that the $j$-th state has been selected as a candidate for replacement and $\alpha$ is given by Eq. (6). Since $j \in \{1, \ldots, N\}$, for the interchangeability we have $N$ equal probabilities (this is the reason of the factor $N$). Replacing the expression of $\alpha$ in Eq. (6), we obtain

$$\begin{aligned}
K(\mathcal{P}_t|\mathcal{P}_{t-1}) &= N\varphi(\mathbf{x}_{0,t-1}) \frac{\frac{\varphi(\mathbf{x}_{j,t-1})}{\pi(\mathbf{x}_{j,t-1})}}{\sum_{i=1}^{N} \frac{\varphi(\mathbf{x}_{i,t-1})}{\pi(\mathbf{x}_{i,t-1})}} \frac{\sum_{i=1}^{N} \frac{\varphi(\mathbf{x}_{i,t-1})}{\pi(\mathbf{x}_{i,t-1})}}{\sum_{i=0}^{N} \frac{\varphi(\mathbf{x}_{i,t-1})}{\pi(\mathbf{x}_{i,t-1})} - \min_{0 \leq i \leq N} \frac{\varphi(\mathbf{x}_{i,t-1})}{\pi(\mathbf{x}_{i,t-1})}}, \\
&= \frac{N}{\pi(\mathbf{x}_{j,t-1})} \frac{\varphi(\mathbf{x}_{0,t-1})\varphi(\mathbf{x}_{j,t-1})}{\sum_{i=0}^{N} \frac{\varphi(\mathbf{x}_{i,t-1})}{\pi(\mathbf{x}_{i,t-1})} - \min_{0 \leq i \leq N} \frac{\varphi(\mathbf{x}_{i,t-1})}{\pi(\mathbf{x}_{i,t-1})}}.
\end{aligned}$$

Now, we can also write

$$\bar{\pi}_g(\mathcal{P}_{t-1})K(\mathcal{P}_t|\mathcal{P}_{t-1}) = \left[\prod_{i=1}^{N} \bar{\pi}(\mathbf{x}_{i,t-1})\right] \frac{N}{\pi(\mathbf{x}_{j,t-1})} = \frac{\varphi(\mathbf{x}_{0,t-1})\varphi(\mathbf{x}_{j,t-1})}{\sum_{i=0}^{N} \frac{\varphi(\mathbf{x}_{i,t-1})}{\pi(\mathbf{x}_{i,t-1})} - \min_{0 \leq i \leq N} \frac{\varphi(\mathbf{x}_{i,t-1})}{\pi(\mathbf{x}_{i,t-1})}},$$

and defining $\gamma(\mathcal{P}_{t-1}, \mathbf{x}_{0,t-1}) = \sum_{i=0}^{N} \frac{\varphi(\mathbf{x}_{i,t-1})}{\pi(\mathbf{x}_{i,t-1})} - \min_{0 \leq i \leq N} \frac{\varphi(\mathbf{x}_{i,t-1})}{\pi(\mathbf{x}_{i,t-1})}$, we have

$$\bar{\pi}_g(\mathcal{P}_{t-1})K(\mathcal{P}_t|\mathcal{P}_{t-1}) = \frac{N}{Z}\left[\prod_{i=1 \neq j}^{N} \bar{\pi}(\mathbf{x}_{i,t-1})\right] \frac{\varphi(\mathbf{x}_{0,t-1})\varphi(\mathbf{x}_{j,t-1})}{\gamma(\mathcal{P}_{t-1}, \mathbf{x}_{0,t-1})},$$

where $Z = \int_{\mathcal{D}} \pi(\mathbf{x})d\mathbf{x}$. This expression above is symmetric w.r.t. $\mathbf{x}_{0,t-1}$ and $\mathbf{x}_{j,t-1}$. Since $\mathcal{P}_{t-1}$ and $\mathcal{P}_t$ differ only in the elements $\mathbf{x}_{0,t-1}$ and $\mathbf{x}_{j,t-1}$ ($\mathcal{P}_{t-1}$ contains $\mathbf{x}_{j,t-1}$ whereas $\mathcal{P}_t$ contains $\mathbf{x}_{0,t-1}$), then $\bar{\pi}_g(\mathcal{P}_{t-1})K(\mathcal{P}_t|\mathcal{P}_{t-1}) = \bar{\pi}_g(\mathcal{P}_t)K(\mathcal{P}_{t-1}|\mathcal{P}_t)$, which is precisely the detailed balance condition.

## E. Ergodicity of the parallel schemes based on multiple candidates

Similarly as in PMC, the parallel Ensemble MCMC (P-EnM) and Multiple Try Metropolis (P-MTM) schemes in Tables 5-6 are based on the particle approximations of the measure of the target. In both cases, $L$ independent samples $\mathbf{z}_1, \ldots, \mathbf{z}_L$ drawn from $\psi(\mathbf{x})$, i.e.,

$$\mathbf{z}_\ell \sim \psi(\mathbf{x}), \tag{27}$$

for $\ell = 1, \ldots, L$. Below, we show that P-EnM and P-MTM yield reversible chains with stationary density the generalized pdf $\bar{\pi}_g$, proving the detailed balance condition is satisfied [4].

### E.1. Parallel Multiple Try Metropolis

In P-MTM, we can define the particle approximation based on the set $\{\mathbf{z}_1, \ldots, \mathbf{z}_L\}$, i.e.,

$$\hat{\pi}^{(L)}(\mathbf{z}) = \sum_{\ell=1}^{L} \beta_\ell \delta(\mathbf{z} - \mathbf{z}_\ell), \tag{28}$$

where the normalized weights $\beta_\ell$'s are given in Eq (10). Note that, the expression above coincides with Eq. (23). Let us also denote as the matrix

$$\mathbf{Z}_{\neg k} = [\mathbf{z}_1, \ldots, \mathbf{z}_{k-1}, \mathbf{z}_{k+1}, \ldots, \mathbf{z}_L],$$

containing all the samples $\mathbf{z}_\ell$'s with the exception of $\mathbf{z}_k$. We denote as $K_n(\mathbf{x}_{n,t}|\mathbf{x}_{n,t-1})$ is the MTM kernel of $n$-th chain, namely, $K_n(\mathbf{z}|\mathbf{x})$ is the probability of the $n$-th chain of jumping from the state $\mathbf{x} = \mathbf{x}_{t-1}$ to $\mathbf{z} = \mathbf{z}_k \in \{\mathbf{z}_1, \ldots, \mathbf{z}_L\}$ (for simplicity, we consider here only the case $\mathbf{z} \neq \mathbf{x}$). Note that all the $\mathbf{z}_\ell$'s are both drawn and resampled independently (see steps 2(a) and 2(b) in Table 6). Thus, the conditional probability $K_n(\mathbf{z}|\mathbf{x})$ can be expressed as

$$K_n(\mathbf{z} = \mathbf{z}_k|\mathbf{x}) = \sum_{\ell=1}^{L} K_n(\mathbf{z}_k|\mathbf{x}, k = \ell),$$

$$= L \int_{\mathcal{D}^{L-1}} \left[ \prod_{\ell=1}^{L} \psi(\mathbf{z}_\ell) \right] \hat{\pi}_{MTM}^{(L)}(\mathbf{z}_k) \ \alpha_n(\mathbf{x}, \mathbf{z}_k|\mathbf{Z}_{\neg k}) \ d\mathbf{Z}_{\neg k}.$$

for $\mathbf{z} \neq \mathbf{x}$, where the function $\alpha_n$ is given in Eq. (11) and we have considered the case $\mathbf{x}$ and $\mathbf{z}$ (the case, $\mathbf{z} = \mathbf{x}$ is straightforward). The factor $L$ is due of the exchangeability among the $L$ random candidates. Thus, we can also write

$$\bar{\pi}(\mathbf{x})K_n(\mathbf{z}_k|\mathbf{x}) =$$

$$L\bar{\pi}(\mathbf{x})\psi(\mathbf{z}_k) \int_{\mathcal{D}^{L-1}} \left[ \prod_{\ell=1;\ell \neq k}^{L} \psi(\mathbf{z}_\ell) \right] \beta_k \alpha_n(\mathbf{x}, \mathbf{z}_k|\mathbf{Z}_{\neg k}) \ d\mathbf{Z}_{\neg k},$$

$$= \frac{L}{Z}\pi(\mathbf{x})\pi(\mathbf{z}_k) \int_{\mathcal{D}^{L-1}} \left[ \prod_{\ell=1;\ell \neq k}^{L} \psi(\mathbf{z}_\ell) \right] \frac{1}{\sum_{\ell=1}^{L} \frac{\pi(\mathbf{z}_\ell)}{\psi(\mathbf{z}_\ell)}} \alpha_n(\mathbf{x}, \mathbf{z}_k|\mathbf{Z}_{\neg k}) \ d\mathbf{Z}_{\neg k},$$

where we have also used the equality $\bar{\pi}(\mathbf{x}) = \frac{1}{Z}\pi(\mathbf{x})$. Replacing

$$\alpha_n(\mathbf{x}, \mathbf{z}_k|\mathbf{Z}_{\neg k}) = \min \left[ 1, \frac{\sum_{\ell=1}^{L} \frac{\pi(\mathbf{z}_\ell)}{\psi(\mathbf{z}_\ell)}}{\sum_{\ell=1}^{L} \frac{\pi(\mathbf{z}_\ell)}{\psi(\mathbf{z}_\ell)} - \frac{\pi(\mathbf{z}_k)}{\psi(\mathbf{z}_k)} + \frac{\pi(\mathbf{x})}{\psi(\mathbf{x})}} \right],$$

in the expression (29) and with some simple rearrangements, we obtain

$$\bar{\pi}(\mathbf{x})K_n(\mathbf{z}_k|\mathbf{x}) = \frac{L}{Z}\pi(\mathbf{x})\pi(\mathbf{z}_k) \int_{\mathcal{D}^{L-1}} \left[ \prod_{\ell=1;\ell \neq k}^{L} \psi(\mathbf{z}_\ell) \right] \min \left[ \frac{1}{\sum_{\ell \neq k} \frac{\pi(\mathbf{z}_\ell)}{\psi(\mathbf{z}_\ell)} + \frac{\pi(\mathbf{z}_k)}{\psi(\mathbf{z}_k)}}, \frac{1}{\sum_{\ell \neq k} \frac{\pi(\mathbf{z}_\ell)}{\psi(\mathbf{z}_\ell)} + \frac{\pi(\mathbf{x})}{\psi(\mathbf{x})}} \right] \ d\mathbf{Z}_{\neg k}.$$

We can observe that, in equation above, we can exchange the position of the variables $\mathbf{x}$ so that $\mathbf{z}_k$ and the expression does not change. So that we can write

$$\bar{\pi}(\mathbf{x})K_n(\mathbf{z}_k|\mathbf{x}) = \bar{\pi}(\mathbf{z}_k)K_n(\mathbf{x}|\mathbf{z}_k), \qquad (29)$$

for all $n = 1, \ldots, N$. The expression above is the so-called *detailed balance condition* [4]: since it holds for all $n$, the complete horizontal MTM process has $\bar{\pi}_g$ as invariant pdf.

### E.1.1. Important observations and Block Independent MTM

First of all, note that with respect to a standard parallel multiple try approach, the novel P-MTM scheme generates only $L$ candidates at each iteration, instead of $NL$ samples. Indeed, P-MTM "recycles" the samples $\mathbf{z}_1, \ldots, \mathbf{z}_L$ from the independent proposal pdf $\psi(\mathbf{x})$, using them in all the $N$ chains. Namely, in P-MTM, at one iteration, the different MTM chains share the same set of tries. However, looking a single chain, each time $L$ new samples are drawn from $\psi(\mathbf{x})$ so that the chain is driven exactly from a standard (valid) MTM kernel. Figures 10(a) and (b) compare graphically the standard parallel MTM approach and the P-MTM scheme (with $N = 2$ chains and $L = 3$ tries). Observe that, in Figure 10(a), 12 new evaluations of the target are needed whereas only 6, in Figure 10(b).

Using the same arguments, the method remains valid if only one resampling step is performed at each iteration, providing one $\mathbf{z}^*$: in this case the same $\mathbf{z}^*$ is tested in the different acceptance tests of the $N$ parallel MTM chains, at the same iteration (exactly as in Table 3 and Fig. 4 for MH kernels). Figure 10(c) shows this case. In order to reduce the possible loss of the diversity, since several chains could jump at the same new state $\mathbf{z}^*$, an alternative strategy can be employed: the Block Independent MTM (BI-MTM) algorithm described in Table 13. Since the proposal $\psi$ is independent and then fixed, before a block of $N$ transitions, we can draw $NL$ tries from $\psi(\mathbf{x})$. Then, we can divide them in $N$ sets $\mathcal{S}_j$, with $j = 1, \ldots, N$ and select one sample from each set, obtaining $\{\mathbf{z}_{k_1}, \ldots, \mathbf{z}_{k_N}\}$ with $\mathbf{z}_{k_j} \in \mathcal{S}_j$. Then, we use $N$ different permutations of $\{\mathbf{z}_{k_1}, \ldots, \mathbf{z}_{k_N}\}$ for performing $N$ iterations of the $N$ parallel chains, providing a better mixing with respect to the case in Figure 10(c). This strategy, i.e., the BI-MTM scheme, is perfectly equivalent to the previous one, shown in Figure 10(c), from a theoretical and computational point of view. BI-MTM is represented graphically in Figure 10(d).

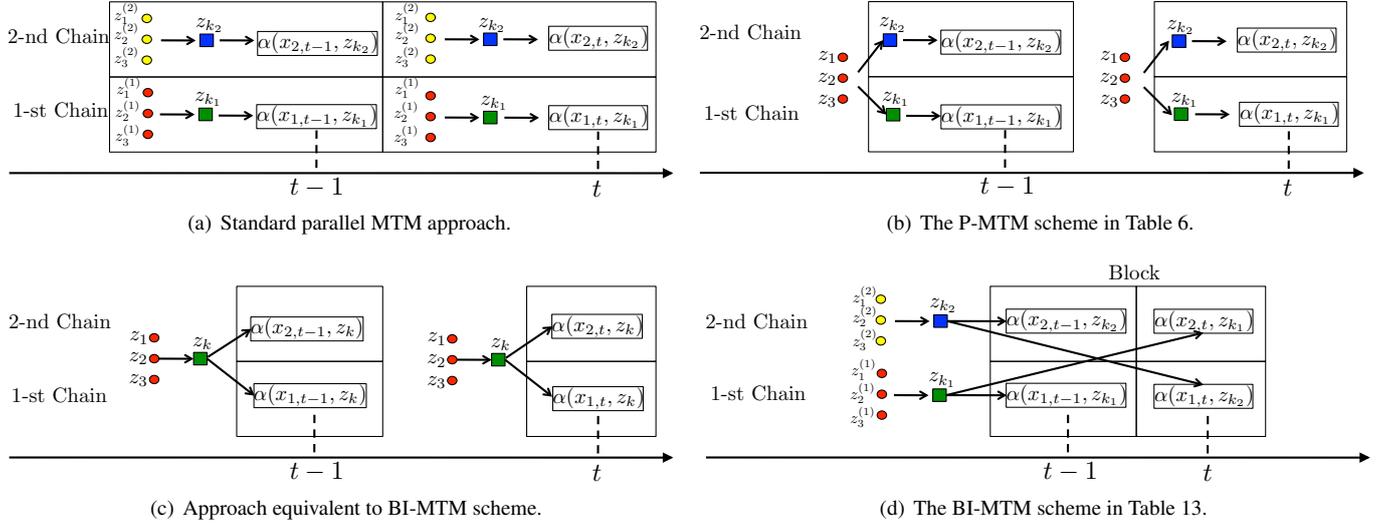

(a) Standard parallel MTM approach.

(b) The P-MTM scheme in Table 6.

(c) Approach equivalent to BI-MTM scheme.

(d) The BI-MTM scheme in Table 13.

Figure 10. A graphical representation of the several parallel MTM schemes with $N = 2$ chains and $L = 3$ tries. The BI-MTM scheme in **(d)** requires only 6 evaluations of the target pdf and 2 multinomial sampling steps considering two iterations, $t - 1$ and $t$.

### E.2. Parallel Ensemble MCMC

Let us consider now the method in Table 5. In this case, the particle approximation is

$$
\begin{aligned}
\hat{\pi}_n^{(L+1)}(\mathbf{z}) &= \sum_{\ell=1}^{L} \alpha_\ell \delta(\mathbf{z} - \mathbf{z}_\ell) + \alpha_{L+1} \delta(\mathbf{z} - \mathbf{x}_{n,t-1}) \\
&= \sum_{\ell=1}^{L+1} \alpha_\ell \delta(\mathbf{z} - \mathbf{z}_\ell),
\end{aligned}
\tag{30}
$$

where $\mathbf{z}_{L+1} = \mathbf{x}_{n,t-1}$. In this case, for a given $n = 1, \ldots, N$, the conditional probability $K_n(\mathbf{z} = \mathbf{z}_k | \mathbf{x})$, where $\mathbf{x} = \mathbf{x}_{n,t-1}$ and $\mathbf{z}_k \in \{\mathbf{z}_1, \ldots, \mathbf{z}_L, \mathbf{z}_{L+1} = \mathbf{x}_{n,t-1}\}$, is given by

$$
\begin{aligned}
K_n(\mathbf{z}_k | \mathbf{x}) &= \sum_{\ell=1}^{L} K_n(\mathbf{z}_k | \mathbf{x}, k = \ell), \\
&= L \int_{\mathcal{D}^{L-1}} \left[ \prod_{\ell=1}^{L} \psi(\mathbf{z}_\ell) \right] \hat{\pi}_n^{(L+1)}(\mathbf{z}_k) \; d\mathbf{Z}_{-k},
\end{aligned}
\tag{31}
$$

Table 13. Block Independent Multiple Try Metropolis (BI-MTM) algorithm for $N$ parallel chains.

---

1. Let $N$ be the total number of parallel MTM chains and $T_H$ be the total number of iterations of each chain, such that $\frac{T_H}{N} \in \mathbb{N}$. Choose a number of tries $L$. sSet $t_0 = mT_V + (m-1)T_H$ if BI-MTM is used within O-MCMC. Otherwise, set $t_0 = 0$.

2. For each block $b = 1, \ldots, B = \frac{T_H}{N}$ do:

   (a) Draw $NL$ i.i.d. candidates $\mathbf{z}_1^{(h)}, \ldots, \mathbf{z}_L^{(h)} \sim \psi(\mathbf{x})$, for $h = 1, \ldots, N$.

   (b) Draw one sample $\mathbf{z}_{k_h}$ from each set $\bar{\mathcal{S}}_h = \{\mathbf{z}_1^{(h)}, \ldots, \mathbf{z}_L^{(h)}\}$ for $h = 1, \ldots, N$, with probability

   $$\beta_\ell^{(h)} = \frac{\frac{\pi(\mathbf{z}_\ell^{(h)})}{\psi(\mathbf{z}_\ell^{(h)})}}{\sum_{\ell=1}^L \frac{\pi(\mathbf{z}_\ell^{(h)})}{\psi(\mathbf{z}_\ell^{(h)})}}.$$

   Thus, finally we have $N$ different samples, $\{\mathbf{z}_{k_1}, \ldots, \mathbf{z}_{k_N}\}$, such that $\mathbf{z}_{k_h} \in \mathcal{S}_h$ for $h = 1, \ldots, N$.

   (c) Create the circular permutations $\mathbf{v}_{n,j} \in \{\mathbf{z}_{k_1}, \ldots, \mathbf{z}_{k_N}\}$ as defined in Eq. (12).

   (d) For $t = (b-1)N + 1 + t_0, \ldots, bN + t_0$ (i.e., exactly $N$ transitions):

       i. Set $j = t - (b-1)N - t_0$ (so that $j = 1, \ldots, N$, in one block).

       ii. For $n = 1, \ldots, N$:

          A. Set $\mathbf{x}_{n,t} = \mathbf{v}_{n,j}$, with probability

   $$\alpha_n(\mathbf{x}_{n,t-1}, \mathbf{v}_{n,j}) = \min\left[1, \frac{\sum_{\ell=1}^L \frac{\pi(\mathbf{z}_\ell^{(j)})}{\psi(\mathbf{z}_\ell^{(j)})}}{\sum_{\ell=1}^L \frac{\pi(\mathbf{z}_\ell^{(j)})}{\psi(\mathbf{z}_\ell^{(j)})} - \frac{\pi(\mathbf{v}_{n,j})}{\psi(\mathbf{v}_{n,j})} + \frac{\pi(\mathbf{x}_{n,t-1})}{\psi(\mathbf{x}_{n,t-1})}}\right].$$

          Otherwise, set $\mathbf{x}_{n,t} = \mathbf{x}_{n,t-1}$.

       iii. Set $\mathcal{P}_t = \{\mathbf{x}_{1,t}, \ldots, \mathbf{x}_{N,t}\}$.

---

for $\mathbf{z} \neq \mathbf{x}$. After some simple rearrangements (similarly in P-MTM) and using the formula of the weights in Eq. (8), we obtain

$$\bar{\pi}(\mathbf{x})K_n(\mathbf{z}_k|\mathbf{x}) = L\bar{\pi}(\mathbf{x})\psi(\mathbf{z}_k) \int_{\mathcal{D}^{L-1}} \left[\prod_{\ell=1,\ell \neq k}^L \psi(\mathbf{z}_\ell)\right] \frac{\frac{\pi(\mathbf{z}_k)}{\psi(\mathbf{z}_k)}}{\sum_{\ell=1}^L \frac{\pi(\mathbf{z}_\ell)}{\psi(\mathbf{z}_\ell)} + \frac{\pi(\mathbf{x})}{\psi(\mathbf{x})}} d\mathbf{Z}_{\neg k},$$

$$= \frac{L}{Z}\pi(\mathbf{x})\pi(\mathbf{z}_k) \int_{\mathcal{D}^{L-1}} \left[\prod_{\ell=1,\ell \neq k}^L \psi(\mathbf{z}_\ell)\right] \frac{1}{\sum_{\ell=1;\ell \neq k}^L \frac{\pi(\mathbf{z}_\ell)}{\psi(\mathbf{z}_\ell)} + \frac{\pi(\mathbf{z}_k)}{\psi(\mathbf{z}_k)} + \frac{\pi(\mathbf{x})}{\psi(\mathbf{x})}} d\mathbf{Z}_{\neg k}.$$

Observing the last equation, we can clearly replace the variable $\mathbf{x}$ with $\mathbf{z}_k$ and vice versa, without changing the expression. Hence, finally we obtain

$$\bar{\pi}(\mathbf{x})K_n(\mathbf{z}_k|\mathbf{x}) = \bar{\pi}(\mathbf{z}_k)K_n(\mathbf{x}|\mathbf{z}_k),$$

for all $n = 1, \ldots, N$, that is the detailed balance condition. For further considerations, see App. E.1.1 above.